\documentclass[12pt]{article}
\pdfoutput=1
\usepackage{jheppub_X}

\usepackage[utf8]{inputenc}

\usepackage{adjustbox}

\usepackage{cleveref}
\crefname{table}{Table}{Tables}
\crefname{equation}{Eq.}{Eqs.}
\crefname{appendix}{App.}{Apps.}
\crefname{section}{Sec.}{Secs.}
\crefname{figure}{Fig.}{Figs.}

\usepackage{setspace}
\usepackage{tikz}
\usetikzlibrary{decorations.markings}
\usepackage{tcolorbox}
\usepackage{bm}
\usepackage{xfrac}
\usepackage{pbox}
\usepackage{graphicx}
\usepackage{bbold}
\usepackage{slashed,braket}
\usepackage{physics}
\usepackage{graphbox}

\allowdisplaybreaks

\makeatletter
\g@addto@macro\bfseries{\boldmath}
\makeatother

\definecolor{colorTC}{rgb}{.2,.7,.2}

\def\hc{\text{h.c.}}
\def\eg{\textit{e.g.}}

\newcommand{\lag}{\ensuremath{\mathcal{L}}}

\newcommand{\s}{\hspace{0.8pt}}

\renewcommand{\d}{\delta}

\newcommand{\g}{\gamma}

\newcommand{\h}{\eta}
\newcommand{\q}{\theta}
\renewcommand{\k}{\kappa}

\newcommand{\p}{\pi}

\newcommand{\si}{\sigma}

\renewcommand{\t}{\tau}

\title{\huge
Non-Decoupling New Particles
}

\author[a]{Ian Banta,}
\author[b]{Timothy Cohen,}
\author[a,c,d]{Nathaniel Craig,}
\author[b]{Xiaochuan Lu,}
\author{\\[5pt] and}
\author[e]{Dave Sutherland}

\affiliation[a]{Department of Physics, University of California, Santa Barbara, CA 93106, USA}
\affiliation[b]{Institute for Fundamental Science, University of Oregon, Eugene, Oregon 97403, USA}
\affiliation[c]{Physics Division, Lawrence Berkeley National Laboratory, Berkeley, CA 94720, USA}
\affiliation[d]{Berkeley Center for Theoretical Physics, University of California, Berkeley, CA 94720, USA}
\affiliation[e]{INFN Sezione di Trieste, via Bonomea 265, 34136 Trieste TS, Italy}

\emailAdd{banta@ucsb.edu}
\emailAdd{tcohen@uoregon.edu}
\emailAdd{ncraig@ucsb.edu}
\emailAdd{xlu@uoregon.edu}
\emailAdd{dsutherl@sissa.it}

\abstract{
We initiate the study of a new class of beyond the Standard Model states that we call ``Loryons.'' They have the defining characteristic of being non-decoupling, in the sense that their physical mass is dominated by a contribution from the vacuum expectation value of the Higgs boson.  The stakes are high:  the discovery of a Loryon would tell us that electroweak symmetry must be non-linearly realized in the effective field theory of the Standard Model. Loryons have their masses bounded from above by perturbative unitarity considerations and thus define a finite parameter space for exploration.  After providing a complete catalog of Loryon representations under mild assumptions, we turn to examining the constraints on the parameter space from Higgs couplings measurements, precision electroweak tests, and direct collider searches. We show that most fermionic candidates are already ruled out (with some notable exceptions), while much of the scalar Loryon parameter space is still wide open for discovery.
}

\begin{document}
\maketitle
\flushbottom
\setcounter{page}{2}

\section{Introduction}

Have we discovered all of the particles that acquire most of their mass from the Higgs? Thanks to decoupling, whether we have discovered all particles that acquire {\it any} of their mass from the Higgs is essentially unknowable with finite experimental precision. But particles acquiring a fixed fraction of their mass from the Higgs are effectively non-decoupling as the strength of their interaction with the Higgs necessarily grows in proportion to their mass. Among these, perhaps the most interesting are particles acquiring {\it the majority} of their mass from the Higgs.\footnote{We will sharpen the notion of ``majority'' in what follows; the detailed criteria varies weakly depending on the quantum numbers of the new particles and the nature of their couplings to the Higgs, but in all cases it roughly corresponds to particles obtaining more than half of their mass from the Higgs.} The low-energy effects of such particles must be described by the $U(1)_{\rm em}$-symmetric Higgs EFT (HEFT) rather than the $SU(2)_L \times U(1)_Y$-symmetric Standard Model EFT (SMEFT). The underlying reason is that the low-energy theory obtained by integrating out such particles does not admit a SMEFT-like expansion around a $SU(2)_L \times U(1)_Y$-preserving point in the EFT field space that converges at our observed vacuum \cite{Cohen:2020xca}. In this sense, particles acquiring most of their mass from the Higgs provide simple, perturbative UV completions of HEFT that cannot be described using SMEFT \cite{Alonso:2015fsp,Alonso:2016oah,Falkowski:2019tft,Cohen:2020xca}. Insofar as their masses are bounded from above  by unitarity considerations to be $\lesssim 4 \pi v$ (where $v$ is the Higgs boson vacuum expectation value), these particles provide a well-defined and entirely finite target for experimental searches.

We follow in the footsteps of Gell-Mann and refer to such particles as {\it Loryons}.\footnote{From {\it Finnegan's Wake}, ``with Pa's new heft...see Loryon the comaleon.'' Note that Loryons whose masses vanish as $v \rightarrow 0$ were referred to as ``Higgs descendents'' in \cite{Cheung:2011aa}. Such particles form a particularly interesting subset of Loryons, as the EFT manifold obtained by integrating them out does not contain an $SU(2)_L \times U(1)_Y$-preserving fixed point. But our interest lies in the broader class of perturbative UV completions of HEFT that cannot be described using SMEFT, for which imposing $m \rightarrow 0$ as $v \rightarrow 0$ is too strong of a requirement.} Most of the fundamental particles in the Standard Model are themselves Loryons, with the exception of the photon and gluon (though the pedantic among us might argue for their inclusion as well, inasmuch as their Higgs-independent masses are not larger than their Higgs dependent-ones, both being zero). This classification is perhaps even more apt when applied to mesons and baryons, whose masses arise only in part from electroweak symmetry breaking; pions, kaons, and $B$ mesons are Loryons, while protons and neutrons are not. Our goal here is to explore the phenomenology of Beyond the Standard Model (BSM) Loryons, with a particular eye towards the parameter space that could yield a future discovery.

The search for BSM Loryons has a long history.  Perhaps the most notable example is the extension of the Standard Model (SM) that includes a chiral fourth generation, the smallest anomaly-free set of entirely chiral fermions that all carry SM charges. Famously, this model is excluded by a combination of unitarity bounds, Higgs coupling measurements, precision electroweak tests, and direct searches~\cite{Djouadi:2012ae, Kuflik:2012ai}. But this leaves the door open for vector-like fermions, scalars, or vectors in various representations of $SU(3)_C \times SU(2)_L \times U(1)_Y$. In the case of vector-like fermions, even though their vector-like masses are allowed by their gauge quantum numbers, it is technically natural for them to receive the majority of their mass from electroweak symmetry breaking since these terms may still violate global symmetries. Aspects of this scenario and its implications for the EFT of the Higgs were recently studied in \cite{Bonnefoy:2020gyh}. For scalars, there are (famously) fewer symmetries available to protect possible mass terms.  So while such BSM examples are still phenomenologically interesting, justifying a small mass that is independent of electroweak symmetry is harder from a symmetry perspective.\footnote{But not impossible -- for instance, a Goldstone boson whose coupling to the Higgs provides the leading violation of its shift symmetry can naturally acquire most of its mass from electroweak symmetry breaking.} In the case of vectors, gauge symmetry provides a natural way of controlling mass terms, but strong constraints exist for models where the Higgs is directly charged under the new gauge group so that it can be primarily responsible for generating the associated gauge boson masses. A more viable option is for the non-zero Higgs vacuum expectation value to induce spontaneous symmetry breaking of the BSM local symmetry via another scalar, in which case the vector bosons can acquire most of their mass from the Higgs without coupling to it via renormalizable operators. Such indirect scenarios were highlighted in \cite{Cheung:2011aa}. For simplicity, in what follows we will focus on scalar and fermionic Loryons obtaining the majority of their mass from a direct coupling to the Higgs. As we will see, there is no shortage of such candidates, but it would also be interesting to further explore the space of Loryon candidates indirectly acquiring the majority of their mass from the Higgs.

To the extent that they remain viable, BSM Loryons provide compelling motivation to use HEFT to parameterize possible deviations in Higgs coupling measurements. Of course, it is reasonable to ask whether one should use an EFT to describe the low-energy effects of Loryons in the first place given that their masses are necessarily bounded to lie below $\sim 4 \pi v$. Here it bears emphasizing that the vast majority of collisions at the LHC involve partonic energies below 1 TeV, and in particular the events most relevant to precision Higgs measurements involve partonic energies much closer to the weak scale. In this regime, EFTs truncated at lower orders in their derivative expansion (as is the case in the practical application of HEFT and SMEFT) are appropriate. While it is possible that some Loryons may be discovered (or excluded) predominantly by direct searches, as we will see, Higgs couplings and other Standard Model precision measurements provide a complementary (and perhaps even the best) probe of Loryon parameter space. Signals in these channels are generally well-described by an EFT, and the first signs of a deviation are likely to be presented in an EFT framework. As such, the existence of Loryons consistent with current data provides a strong motivation to use HEFT as the BSM parameterization when performing future searches for Higgs coupling deviations.

This paper is organized as follows: In \cref{sec:candidates}, we enumerate scalar and fermionic Loryons obtaining the majority of their mass from direct coupling to the Higgs. For simplicity, and to minimize constraints from precision electroweak measurements, we only study candidates that can be expressed as multiplets of the approximate $SU(2)_L \times SU(2)_R$ custodial symmetry of the SM; we allow for any custodially symmetric renormalizable couplings to the Higgs. We present a sharp criterion for determining when the local EFT obtained by integrating out these custodial irreducible representations (irreps) must be HEFT (in that it cannot be written in terms of a convergent SMEFT at our vacuum) and use this to define the parameter space of interest for Loryons. In these cases, the mass scale of weakly-coupled Loryons is bounded from above by perturbative unitarity considerations. While it is certainly possible for Loryon couplings to exceed these bounds, we use them to determine the regime in which our perturbative calculations remain under control. In \cref{sec:higgs}, we consider constraints on the Loryon candidates coming from Higgs coupling measurements, most notably LHC bounds on $h \rightarrow \gamma \gamma$, $h \rightarrow gg$, $h \rightarrow$ invisible, and $h \rightarrow$ untagged. We then turn to consider precision electroweak limits in \cref{sec:PEWK}, which largely stem from bounds on the $S$ parameter given our imposition of custodial symmetry. In \cref{sec:direct}, we determine the current state of direct limits on the Loryon candidates that remain viable after imposing perturbative unitarity bounds, Higgs couplings, and precision electroweak measurements are taken into account. We summarize the surviving parameter space of Loryons in \cref{sec:viable}, finding a number of compelling candidates that are consistent with all known data, thereby providing a sharp target for future searches. We briefly sketch the future prospects for the HL-LHC in \cref{sec:future}, focusing on projected improvements in Higgs coupling measurements; both $h \rightarrow Z \gamma$ and $h$ pair production are expected to provide qualitatively new sensitivity. We present our conclusions in \cref{sec:conclusion}. The calculation of the mass spectrum and one-loop contribution to the Higgs wave function renormalization is relegated to the Appendix.

\section{Loryon Catalog}
\label{sec:candidates}

Our starting point is to enumerate the BSM Loryons that have a possibility of being phenomenologically viable.  Our first goal will be to specify their SM quantum numbers and to understand the implications for the allowed mass terms and couplings to the Higgs field in \cref{subsec:reps}.  We will then discuss the connection to HEFT in \cref{subsec:criteria} by specifying the conditions under which integrating out a Loryon requires matching the resulting theory onto HEFT. In these cases, the mass of the Loryons are bounded from above by perturbative unitarity considerations, which we will explore in \cref{subsec:unitarity}.

\subsection{Representations and Mass Spectrum}
\label{subsec:reps}

New physics that badly violates the approximate $SU(2)_L \times SU(2)_R$ custodial symmetry of the SM is strongly constrained by precision electroweak measurements.\footnote{In fact, the lack of evidence for additional custodial symmetry violation is often taken as a reason for preferring to interpret experimental results using SMEFT over HEFT.  We caution that this is not a valid argument since both EFT frameworks admit custodially symmetric limits.} In order to focus on the candidates that are most compatible with current data, we will restrict our attention to Loryons that preserve custodial symmetry. There are, of course, viable Loryon candidates that violate custodial symmetry, albeit with parameter spaces more tightly constrained by precision electroweak data; we leave the study of this broader class of candidates to future work.

Given this assumption, we will refer to the representation of the BSM Loryons in two ways, depending on the context.  One useful approach is to denote the representation under the SM gauge groups; this will be written using the notation $(C,L)_Y$ for the charges under $SU(3)_C$, $SU(2)_L$, and $U(1)_Y$.  The other notation is to specify the representation under custodial symmetry; this will be written as $[L,R]_Y$, where we suppress the color information, and use the integers $L/R$ to denote the dimensions of the representations under $SU(2)_{L/R}$. In this notation, $Y$ denotes the hypercharge of the states \emph{on top of} the eigenvalues of the $SU(2)_R$ generator $T_R^3$, so the custodial representation $[L,R]_Y$ contains SM fields $(C,L)_{-(R-1)/2+Y}$ to $(C,L)_{(R-1)/2+Y}$ in steps of unit hypercharge. We require a priori that all BSM Loryons satisfy the following three conditions:
\begin{itemize}
\item The color singlets have integer electromagnetic charges.
\item Those possessing electromagnetic charges can promptly decay.
\item Fermionic Loryons are introduced in pairs such that one can write a custodial singlet Yukawa term.
\end{itemize}
The prompt decay constraint is taken to be $c\tau\lesssim1$ mm. A rough calculation taking a benchmark suppression scale $\Lambda\gtrsim5$ TeV gives a conservative bound $\dim\le 9$ for the decay operator, which is only logarithmically sensitive to the chosen value of $\Lambda$. This constrains the allowed color, electroweak, and hypercharge representations of BSM Loryons. In \cref{tab:scalarReps} and \cref{tab:fermionReps}, we enumerate the custodial irreps considered, list their maximum allowed hypercharge, and give the labeling of the SM representations following the conventions of~\cite{deBlas:2017xtg}. We do not explicitly enumerate the possible color representations; virtually all of these will be easily ruled out by constraints on the Higgs coupling to gluons parameterized by $\kappa_g$ (see \cref{subsec:hgg}).

\vspace{10pt}
\noindent\textbf{Scalars:} We start with the case of scalar Loryons, listing their SM and custodial representations in \cref{tab:scalarReps}. For each custodial irrep $[L,R]_Y$, we define a $L\times R$ matrix field $\Phi$ that transforms as $\Phi \to U_L \Phi U_R^\dagger$ under the chosen irrep $U_{L/R}$ of $SU(2)_{L/R}$. The custodial irrep is real iff $L+R$ is even and $Y=0$. We can write an explicit mass term for $\Phi$
\begin{equation}
\lag\supset-\frac{m_\text{ex}^2}{2^\rho}\, \tr\left( \Phi^\dagger\Phi \right) \,,
\label{eq:ScalarMassExplicit}
\end{equation}
where $\rho=0\,(1)$ for a complex (real) representation.

\begin{table}[t]
\renewcommand{\arraystretch}{1.6}
\setlength{\tabcolsep}{0.7em}
\setlength{\arrayrulewidth}{1.2pt}
\centering
\begin{tabular}{  c | c | c | c | c | c | c | c  }
\multicolumn{8}{c}{\textsc{Scalars}}\\[5pt]
SM Reps & $(1,1)_Y$ & $(1,2)_Y$ & $(1,3)_Y$ & $(1,4)_Y$ & $(1,L)_Y$ & $(3,1)_Y$ & $(3,2)_Y$ \\
\hline
Field & $S_Y$ & $\Phi_{2Y}$ & $\Xi_Y$ & $\Theta_{2Y}$ & $X_{L,Y}$ & $\omega_{|3Y|}$ & $\Pi_{|6Y|}$
\end{tabular}\\[13pt]
\begin{tabular}{  c | c | c | c | c | c | c | c | c  }
 & $R=1$ & 2 & 3 & 4 & 5 & 6 & 7 & 8 \\
\hline
$L=1$ & $|Y_{max}|=3$ & $\frac52$ & 2 & $\frac32$ & 1 & $\frac12$ & 0 & $\times$ \\
\hline
2 & $\frac72$ & 4 & $\frac72$ & 4 & $\frac92$ & 5 & $\frac{11}{2}$ & 5 \\
\hline
3 & 3 & $\frac72$ & 4 & $\frac92$ & 4 & $\frac92$ & 5 & $\frac{11}{2}$ \\
\hline
4 & $\frac72$ & 3 & $\frac72$ & 4 & $\frac92$ & 5 & $\frac{11}{2}$ & 5 \\
\hline
5 & 3 & $\frac72$ & 4 & $\frac72$ & 4 & $\frac92$ & 5 & $\frac{11}{2}$ \\
\hline
6 & $\frac52$ & 3 & $\frac72$ & 4 & $\frac92$ & 5 & $\frac92$ & 4 \\
\hline
7 & 3 & $\frac72$ & 3 & $\frac52$ & 2 & $\frac32$ & 1 & $\frac12$ \\
\hline
8 & $\frac32$ & 1 & $\frac12$ & 0 & $\times$ & $\times$ & $\times$ & $\times$
\end{tabular}\\[13pt]
\caption{The representations and corresponding fields for the scalar BSM Loryons considered in this work.  We express ``SM charges'' as  $(C,L)_Y$ and ``custodial charges'' as $[L,R]_Y$; the custodial representation $[L,R]_Y$ contains SM fields $(C,L)_{-(R-1)/2+Y}$ to $(C,L)_{(R-1)/2+Y}$ in steps of unit hypercharge. Hypercharges $Y$ are restricted so that any new charged particles can promptly decay.}
\label{tab:scalarReps}
\end{table}

Arranging the components of the Higgs doublet $(\phi_+,\phi_0)^T$ into the $[2,2]_0$ custodial representation
\begin{equation}
H = \mqty(\phi_0^* & \phi_+ \\ - \phi_- & \phi_0)
\,\, \xrightarrow{\text{ unitary gauge }}\,\,
\frac{v+h}{\sqrt{2}} \mqty(1 & 0 \\ 0 & 1) \,,
\label{eq:HiggsMatrixUnitaryGauge}
\end{equation}
we can also write down a Higgs portal interaction
\begin{equation}
\lag\supset-\frac{\lambda_{h\Phi}}{2^\rho}\, \tr\big(\Phi^\dagger\Phi\big)\, \frac12\tr\big(H^\dagger H\big) \,,
  \label{eq:ScalarMassUniv}
\end{equation}
which provides a contribution $\lambda_{h\Phi}v^2/2$ to the mass for all the components of $\Phi$. We will be interested in the Loryon parameter space where the BSM state gets the majority of its mass from electroweak symmetry breaking.  To this end, it is convenient to define the dimensionless quantity $\lambda_\text{ex} \equiv 2 m_\text{ex}^2 / v^2$. Thus the explicit mass term gives a mass-squared $\lambda_\text{ex}v^2/2$ to each component of $\Phi$ (though we emphasize that the explicit masses are independent of the Higgs vev).

Additionally, for the representations that are charged under both $SU(2)_L$ and $SU(2)_R$, there is another contraction for the quartic term:
\begin{equation}
\lag \supset -\frac{\lambda_{h\Phi}'}{2^\rho}\, 2\tr\big(\Phi^\dagger T_{L}^a\Phi T_{R}^{\dot a}\big)\, 2\tr\big(H^\dagger T_2^aH T_2^{\dot a}\big) \,,
\label{eq:ScalarMassSplit}
\end{equation}
involving Hermitian $SU(2)$ generators $T_\text{dim(irrep)}^a$, indexed by $a, \dot a=1,2,3$. In our notation, these generators are canonically normalized
\begin{equation}
\tr\left( T_\text{dim}^a T_\text{dim}^b \right) = \delta^{ab}\, \frac13\, \dim\, C_2(\dim) \,,
\end{equation}
with the Casimir (note that $\dim = 2j+1$)
\begin{equation}
C_2(\dim) = j(j+1) = \frac14 (\dim + 1)(\dim -1) \,.
\end{equation}
After electroweak symmetry breaking, the interaction in \cref{eq:ScalarMassSplit} leads to a mass splitting among the components of $\Phi$. The remaining degeneracies in the mass spectrum arrange into irreps of the unbroken diagonal subgroup $SU(2)_V \subset SU(2)_L \times SU(2)_R$. Explicitly, one can collect the $L\times R$ components of the matrix $\Phi$ into a direct sum of $V$-dimensional vectors $\phi_V$ that are $SU(2)_V$ representations:
\begin{equation}
\Phi \to \underset{V\in\mathcal{V}}{\oplus}\, \phi_V \,,
\label{eqn:ScalarDecomp}
\end{equation}
with
\begin{equation}
\mathcal{V} = \Big\{ L+R-1\;,\; L+R-3\;,\; \cdots \;,\; |L-R|+1 \Big\} \,.
\end{equation}

As a concrete example to illustrate \cref{eqn:ScalarDecomp}, we consider the complex custodial irrep $\Phi\sim[2,3]_{-1/2}$. It has six components and decomposes into a quadruplet and a doublet under $SU(2)_V$:
\begin{equation}\renewcommand\arraystretch{1.2}
\Phi = \mqty( \Phi_{11} & \Phi_{12} & \Phi_{13} \\ \Phi_{21} & \Phi_{22} & \Phi_{23} ) \to \phi_4 \oplus \phi_2 \,.
\label{eqn:PhiExample}
\end{equation}
Working with the $T_R^3$-spin basis, namely each column of the matrix $\Phi$ above has a definite $T_R^3$-spin of $SU(2)_R$ (or equivalently of $SU(2)_V$), respectively $-1, 0, 1$ in order, we can write out each $SU(2)_V$ irrep $\phi_V$ from the $\Phi$ elements using the Clebsch-Gordan coefficients
\begin{equation}\renewcommand\arraystretch{1.5}
\phi_4 = \mqty( \Phi_{13} \\ \frac{1}{\sqrt{3}}\left( -\sqrt{2}\, \Phi_{12}+\Phi_{23} \right) \\ \frac{1}{\sqrt{3}}\left( \Phi_{11} - \sqrt{2}\, \Phi_{22} \right) \\ \Phi_{21} ) \,,\qquad
\phi_2 = \mqty( \frac{1}{\sqrt{3}}\left( \Phi_{12} + \sqrt{2}\, \Phi_{23} \right) \\ \frac{1}{\sqrt{3}}\left( -\sqrt{2}\, \Phi_{11}-\Phi_{22} \right) ) \,.
\label{eqn:CG23}
\end{equation}
Then the interaction in \cref{eq:ScalarMassSplit} reads (in the unitary gauge)
\begin{equation}
\lag \supset -\lambda_{h\Phi}'\, 2\tr\big(\Phi^\dagger T_{L}^a\Phi T_{R}^a \big)\, \frac12 (v+h)^2 = -\frac12 \lambda_{h\Phi}' (v+h)^2\, \left( - \phi_4^\dagger \phi_4 + 2 \phi_2^\dagger \phi_2 \right) \,.
\end{equation}

General cases are worked out systematically in \cref{app:quadlags}. In the end, we obtain the decomposition of \cref{eq:ScalarMassSplit} as
\begin{equation}
\lag \supset -\frac{1}{2^\rho}\, \frac12 \lambda_{h\Phi}' (v+h)^2\, \sum_{V \in \mathcal{V}} \phi_V^\dagger \Big[C_2(L)+C_2(R)-C_2(V)\Big] \phi_V \,.
\end{equation}
This leads us to a convenient form of the quadratic piece of the Lagrangian for an $[L,R]_Y$ scalar Loryon:
\begin{equation}
\mathcal{L}_\text{quad} = - \frac{1}{2^\rho} \sum_{V \in \mathcal{V}} \phi_V^\dagger \left[  D^2 + \frac12 \lambda_\text{ex} v^2 + \frac12 \lambda_V (v+h)^2 \right] \phi_V \,,
\label{eqn:ScalarLag}
\end{equation}
where we have introduced the notation
\begin{equation}
\lambda_V \equiv \lambda_{h\Phi} + \lambda^\prime_{h\Phi} \Big[ C_2(L) + C_2(R) -C_2(V) \Big] \,.
\end{equation}
Note in particular that the mass spectrum of the scalar Loryon is
\begin{equation}
m_V^2 = m_\text{ex}^2 + \frac12 \lambda_V v^2 = \frac12 v^2 \left( \lambda_\text{ex} + \lambda_V \right) \,,\qquad  \forall\; V\in\mathcal{V} \,.
\label{eqn:ScalarMasses}
\end{equation}
Here and henceforth we assume $\lambda_{\text{ex}}, \lambda_{\text{ex}} + \lambda_V \geq0$ in order to ensure stability of the $\Phi=0$ vacuum for any background value of the Higgs.

We will find it useful to scale the mass splitting by the mass common to all the components. To this end, we define
\begin{equation}
r_\text{split}\equiv\frac{\lambda^\prime_{h\Phi}}{\lambda_\text{ex}+\lambda_{h\Phi}}\,.
\label{eqn:rsplit}
\end{equation}

In principle, some scalar BSM Loryons could also couple linearly to various powers of $H$, leading to Higgs-Loryon mixing after electroweak symmetry breaking. We assume these couplings are small, consistent with an approximate discrete symmetry acting on the BSM scalars. We further assume that the Higgs-independent potential for the new scalars is such that they are stabilized at the origin. Under these assumptions, mixing between new scalars and components of the Higgs is a subleading effect on low-energy physics.

\vspace{10pt}
\noindent\textbf{Fermions:} For fermionic Loryons, we consider vector-like fermions whose SM and custodial representations are summarized in \cref{tab:fermionReps}. As in the scalar case, for each custodial irrep $[L,R]_Y$, we define an $L\times R$ matrix field $\Psi$ that transforms as $\Psi \to U_L \Psi U_R^\dagger$. Each element in $\Psi$ is a Dirac field that contains both a left-handed Weyl fermion and a right-handed Weyl fermion. We can then write down an explicit mass term:
\begin{equation}
\lag\supset- M_\text{ex}\, \tr\left(\s \overline\Psi \Psi \right) \,.
\label{eq:Mvl}
\end{equation}
As before, it is convenient to define $y_\text{ex} \equiv \sqrt{2} M_\text{ex}/v$. This facilitates the comparison between the fraction of the fermions' masses that is independent of electroweak symmetry breaking and the fraction that arises from electroweak symmetry breaking. The latter comes from the Yukawa interactions that can be schematically written as
\begin{equation}
\lag\supset-y_{12}\; \overline{\Psi}_{1} \cdot H \cdot \Psi_2 + \hc \,.
\label{eq:Yukawa}
\end{equation}
The representations $[L_1,R_1]_Y$ and $[L_2,R_2]_Y$ (of $\Psi_1$ and $\Psi_2$ respectively) are chosen such that contracting their indices with the $[2,2]_0$ Higgs representation (as schematically denoted by the dot products above) can yield a custodial singlet. This enforces the equality of the hypercharges of two representations and means that $L_1=L_2\pm1$ and $R_1=R_2\pm1$. Henceforth we refer to such a pairing as $[L_1,R_1]_Y \oplus [L_2,R_2]_Y$.

\begin{table}[t!]
\renewcommand{\arraystretch}{1.6}
\setlength{\tabcolsep}{0.7em}
\setlength{\arrayrulewidth}{1.2pt}
\centering
\begin{tabular}{  c | c | c | c | c | c  }
\multicolumn{6}{c}{\textsc{Vector-like Fermions}}\\[5pt]
SM Reps & $(1,1)_Y$ & $(1,2)_Y$ & $(1,3)_Y$ & $(1,L)_Y$ & $(3,1)_Y$ \\
\hline
Field & $E_Y$ & $\Delta_{2Y}$ & $\Sigma_Y$ & $K_{L,Y}$ & $P_{|3Y|}$
\end{tabular}\\[13pt]
\begin{tabular}{  c | c | c | c | c | c | c | c | c  }
 & $R=1$ & 2 & 3 & 4 & 5 & 6 & 7 & 8 \\
\hline
$L=1$ & $|Y_{max}|=3$ & $\frac52$ & 2 & $\frac32$ & 1 & $\frac12$ & 0 & $\times$ \\
\hline
2 & $\frac72$ & 3 & $\frac72$ & 4 & $\frac92$ & 5 & $\frac{11}{2}$ & 5 \\
\hline
3 & 3 & $\frac72$ & 3 & $\frac72$ & 4 & $\frac92$ & 5 & $\frac{11}{2}$ \\
\hline
4 & $\frac52$ & 3 & $\frac72$ & 4 & $\frac72$ & 4 & $\frac92$ & 5 \\
\hline
5 & 3 & $\frac52$ & 3 & $\frac72$ & 4 & $\frac92$ & 5 & $\frac92$ \\
\hline
6 & $\frac52$ & 3 & $\frac72$ & 3 & $\frac52$ & 2 & $\frac32$ & 1 \\
\hline
7 & 2 & $\frac32$ & 1 & $\frac12$ & 0 & $\times$ & $\times$ & $\times$ \\
\hline
8 & $\times$ & $\times$ & $\times$ & $\times$ & $\times$ & $\times$ & $\times$ & $\times$
\end{tabular}\\[13pt]
\caption{The representations and corresponding fields for the vector-like fermionic BSM Loryons considered in this work.  We express ``SM charges'' as  $(C,L)_Y$ and ``custodial charges'' as $[L,R]_Y$; the custodial representation $[L,R]_Y$ contains SM fields $(C,L)_{-(R-1)/2+Y}$ to $(C,L)_{(R-1)/2+Y}$ in steps of unit hypercharge. Hypercharges $Y$ are restricted so that any new charged particles can promptly decay and so that the Yukawa terms with the Higgs are gauge singlets.}
\label{tab:fermionReps}
\end{table}

Upon electroweak symmetry breaking, the Yukawa interaction in \cref{eq:Yukawa} leads to mass splitting among the components of $\Psi_1$ and $\Psi_2$. To keep track of this, we decompose each of them into their respective irreps under the diagonal subgroup $SU(2)_V \subset SU(2)_L \times SU(2)_R$, similar to the scalar case in \cref{eqn:ScalarDecomp}:
\begin{equation}
\Psi_1 \to \underset{V_1\in\mathcal{V}_1}{\oplus}\, \psi_{1,V_1} \,,\qquad
\Psi_2 \to \underset{V_2\in\mathcal{V}_2}{\oplus}\, \psi_{2,V_2} \,,
\label{eqn:PsiDecompose}
\end{equation}
with
\begin{subequations}\label{eqn:V1V2}
\begin{align}
\mathcal{V}_1 &= \Big\{ L_1+R_1-1\;,\; L_1+R_1-3\;,\; \cdots \;,\; |L_1-R_1|+1 \Big\} \,, \\[5pt]
\mathcal{V}_2 &= \Big\{ L_2+R_2-1\;,\; L_2+R_2-3\;,\; \cdots \;,\; |L_2-R_2|+1 \Big\} \,.
\end{align}
\end{subequations}
In the overlap of these two sets $V=V_1=V_2$, the two fermions $\psi_{1,V_1}$ and $\psi_{2,V_2}$ get mass mixings through the interaction in \cref{eq:Yukawa}.

As a concrete example to illustrate this mass mixing, we consider the pair of custodial irreps $\Psi_1\sim [2,3]_{-1/2}$ and $\Psi_2\sim [1,2]_{-1/2}$. The contraction of indices in \cref{eq:Yukawa} can be written out as
\begin{equation}
\lag \supset - y_{12}\, \overline{\Psi}_{1, \alpha \dot{a}}\, H_{\alpha \dot\alpha}\, \frac{1}{\sqrt{2}} \sigma_{\dot\gamma\dot\alpha}^{\dot{a}} \epsilon_{\dot\gamma\dot\beta} \, \Psi_{2,1\dot\beta} + \hc \,, \label{eqn:yukawa2312}
\end{equation}
where undotted (dotted) indices are for $SU(2)_L$ ($SU(2)_R$) and Greek (Latin) indices are used to denote fundamental (adjoint) representations. Upon electroweak symmetry breaking, we have $H_{\alpha \dot\alpha} = \frac{1}{\sqrt{2}}(v+h)\delta_{\alpha\dot\alpha}$ and hence (dropping dots on indices)
\begin{equation}
\lag \supset - \frac12 \, y_{12}\, \overline{\Psi}_{1, \alpha a}\, \sigma_{\gamma\alpha}^{a} \epsilon_{\gamma\beta} \, \Psi_{2,1\dot\beta} + \hc = - \frac{1}{\sqrt{2}}\,\sqrt{\frac32} \, y_{12}(v+h)\, \overline{\psi}_{1,2}\, \psi_{2,2} + \hc \,,
\end{equation}
from which we can make the identification
\begin{equation}
\left(\psi_{1,2}\right)_\beta = \frac{1}{\sqrt{3}}\, \sigma_{\alpha\beta}^a \Psi_{1,\alpha a} \,,\qquad
\left(\psi_{2,2}\right)_\beta = \epsilon_{\beta \alpha} \Psi_{2,1\alpha} \,.
\label{eqn:psi12}
\end{equation}
As expected from \cref{eqn:V1V2}, $\Psi_2$ in this example has a single $SU(2)_V$ irrep with $V_2\in\{2\}$. On the other hand, $\Psi_1$ has two $SU(2)_V$ irreps with $V_1\in\{4,2\}$. The interaction in \cref{eq:Yukawa} gives a mass mixing between the components in the overlap of the two sets $V_1=V_2=2$.\footnote{The expression for $\psi_{1,2}$ in \cref{eqn:psi12} appears to be different from its scalar counterpart (namely $\phi_2$ in \cref{eqn:CG23}). This is because in \cref{eqn:yukawa2312} we used the adjoint basis for the columns of $\Psi_1$, as opposed to the $T_R^3$-spin basis used for $\Phi$ in \cref{eqn:PhiExample}. Carrying out the basis change, one can verify that they are the same linear combination.}

The general cases of a pair of custodial irreps $[L_1,R_1]_Y\oplus[L_2,R_2]_Y$ have been worked out systematically in \cref{app:quadlags}. This leads us to the following general form of the Lagrangian that is quadratic in the fermions
\begin{align}
\lag_\text{quad} &= \sum_{V \in \mathcal{V}_1 - \mathcal{V}_2} \overline{\psi}_{1,V} \left( i\slashed{D} - M_1 \right) \psi_{1,V}
 + \sum_{V \in \mathcal{V}_2 - \mathcal{V}_1} \overline{\psi}_{2,V} \left( i\slashed{D} - M_2 \right) \psi_{2,V} \notag\\[5pt]
& + \sum_{V \in \mathcal{V}_1 \cap \mathcal{V}_2} \mqty( \overline{\psi}_{1,V} & \overline{\psi}_{2,V} ) \left[ i\slashed{D} - \mqty( M_1 & \frac{1}{\sqrt{2}} y_V (v+h) \\ \frac{1}{\sqrt{2}} y_V^* (v+h) & M_2 ) \right] \mqty(\psi_{1,V} \\ \psi_{2,V}) \,,
\label{eqn:FermionLag}
\end{align}
with
\begin{equation}
  y_V = (-1)^{j_1+r_1+l_2+\frac12}\, y_{12}  \times \sqrt{L_1 R_1} \times
\left\{ \begin{matrix}
  r_2 & l_2 & j_1 \\ l_1 & r_1 & \frac12
\end{matrix}
\right\} \, ,
\end{equation}
where $2l_i+1=L_i,2r_i+1=R_i,2j_1+1=V$, and the quantity in brackets is a Wigner 6j symbol. In the analysis that follows, we will take the vector-like masses of the two custodial irreps to be equal $M_1=M_2=M_\text{ex}$ for simplicity; we have checked that this assumption has a minimal impact on our conclusions. Under this choice, the mass spectrum of the fermionic Loryon is
\begin{subequations}\label{eqn:FermionMasses}
\begin{alignat}{2}
M_V &= M_\text{ex} = \frac{v}{\sqrt{2}}\, y_\text{ex} \,,\qquad&
V &\in \mathcal{V}_1 - \mathcal{V}_2 \;\;\text{or}\;\; \mathcal{V}_2 - \mathcal{V}_1 \,, \\[6pt]
M_{\pm V} &= M_\text{ex} \pm \frac{v}{\sqrt{2}} \left|y_V\right| = \frac{v}{\sqrt{2}} \left( y_\text{ex} \pm \left|y_V\right| \right) \,,\qquad&
V &\in \mathcal{V}_1 \cap \mathcal{V}_2 \,.
\end{alignat}
\end{subequations}
Note that for each rep in the second line, $V \in \mathcal{V}_1 \cap \mathcal{V}_2$, the fermion masses come in pairs. When $\left|y_V\right| > y_\text{ex}$, one of the eigenvalues become negative. For this eigenstate of the Dirac fermion, one could flip the relative sign between its two chiral components to absorb the negative sign, so the physical mass of the particle is still positive (at the expense of introducing additional signs into the interactions).

\subsection{Criteria for HEFT}
\label{subsec:criteria}

Given our finite list of BSM Loryons, this section will delineate the regions of Loryon parameter space that require HEFT as the EFT description that emerges at low energies. In principle, the Loryons can give tree- and loop-level effects. The loop-level effects are an irreducible consequence of coupling to the Higgs. Although the tree-level effects would give more striking signatures, to highlight the irreducible effects we assume there is an approximate $\mathbb{Z}_2$ symmetry acting on the BSM Loryon fields, which is respected by all their interactions with the Higgs discussed above; see \cref{eq:ScalarMassSplit,eq:ScalarMassUniv,eq:ScalarMassExplicit} for scalars and \cref{eq:Mvl,eq:Yukawa} for fermions. Close to the perturbative unitarity bound on the $\mathbb{Z}_2$ symmetric couplings, we expect the loop-level effects to be appreciable within the regime of validity of the EFT. As we will see, indirect constraints from Higgs properties are relevant in the parameter space of interest.

In order to compute the leading matching contributions, we utilize functional methods (see \cite{Cohen:2020xca}) to derive the effective (scalar sector) Lagrangian that results if the UV theory includes a scalar Loryon (including all orders in the Higgs field $H$):
\begin{equation}
\mathcal{L}_\text{eff} \supset \frac{1}{2^\rho (4 \pi)^2} \sum_{V\in\mathcal{V}} V \left\{ \frac{m_V^4(H)}{2} \left[ \ln \frac{\mu^2}{ m_V^2(H) } + \frac32 \right] + \frac{\lambda_V^2}{6 m_V^2(H)} \frac{\left[ \partial |H|^2 \right]^2 }{2} + \mathcal{O}\big(\partial^4\big) \right\} \,,
\label{eq:ScalarOneLoopRepeat}
\end{equation}
for an arbitrary custodial irrep $\Phi$; this result is derived in the appendix, see~\cref{eq:ScalarOneLoop}. The appendix also includes the analogous calculation for fermions, see \cref{eq:FermionOneLoop,eq:MForFermionOneLoop}.

Noting that the (non-derivative) dependence on $H$ in~\cref{eq:ScalarOneLoopRepeat} is captured by the effective mass
\begin{equation}
m^2_V(H) = \frac12 \lambda_\text{ex} v^2 + \lambda_V |H|^2 \, ,
\end{equation}
we can then obtain a SMEFT description of \cref{eq:ScalarOneLoopRepeat} by expanding about $H=0$ in powers of $\lambda_V |H|^2 /\left(\frac12 \lambda_\text{ex} v^2 \right)$.
However, the SMEFT description is only useful for predictions of low energy observables if it converges when evaluated about the electroweak breaking vacuum, $|H|=\frac{1}{\sqrt{2}} v$. This requires that\footnote{
  \cref{eq:ScalarOneLoopRepeat} shows a non-analyticity whenever $m_V^2(H) = 0$, which corresponds to a non-analyticity in the complex plane of $|H|$ at $\Im |H| = \pm \frac{\lambda_\text{ex}}{\lambda_V} \frac{v}{\sqrt{2}}$ that limits the radius of convergence of the SMEFT expansion, see \cite{Cohen:2021ucp}.
}
\begin{equation}
\lambda_V < \lambda_\text{ex} \,,\qquad   \forall\; V \in \mathcal{V} \,.
\end{equation}
If this condition is not satisfied, then one is forced to use the HEFT description; see~\cite{Cohen:2020xca}. We will find later that it is useful to introduce a parameter $f_V$:
\begin{equation}
f_V \equiv \frac{\lambda_V}{ \lambda_\text{ex} + \lambda_V } \,,
\label{eqn:fScalar}
\end{equation}
which corresponds to the fraction of a scalar Loryon's physical mass-squared that results from its interactions with the Higgs (see \cref{eqn:ScalarMasses}). Note that this parameter will generically differ among the states $\phi_V$ within a given custodial representation. The criterion for necessarily matching onto HEFT then becomes
\begin{equation}
f_\text{max} \ge \frac12 \,,
\end{equation}
where we have defined $f_\text{max}\equiv\max_{V \in \mathcal{V}} f_V$ as a shorthand for the $f_V$ value of the heaviest scalar state.

A similar story holds for fermionic Loryons.  In this case, the all-orders-in-$H$ effective Lagrangian is given in \cref{eq:FermionOneLoop}.  Its terms are simple functions of the (signed) Higgs-dependent masses of the states
\begin{equation}
M_{\pm V}(H) = M_\text{ex} \pm |y_V| |H| \,,
\end{equation}
and some of them become non-analytic when $M_{\pm V}(H)=0$.\footnote{Despite appearances the effective Lagrangian is well behaved when $M_{+V} = M_{-V}$.}  Then the SMEFT expansion of the effective Lagrangian converges at our vacuum if
\begin{equation}
\left| y_V \right| < y_\text{ex} \,,\qquad   \forall\; V \in \mathcal{V}_1 \cap \mathcal{V}_2 \,.
\end{equation}
Therefore, in terms of the fraction of mass that fermionic Loryons get from interacting with the Higgs (see \cref{eqn:FermionMasses})
\begin{equation}
f_{\pm V} \equiv \frac{\pm\, |y_V| }{y_\text{ex} \pm |y_V|} \,,
\label{eqn:fFermion}
\end{equation}
the criterion for necessarily matching onto HEFT is
\begin{equation}
f_{\max} \ge \frac12 \,,
\end{equation}
where we have defined the shorthand $f_{\max}\equiv\max_{V\in\mathcal{V}_1\cap\mathcal{V}_2} f_{+V}$.

\subsection{Mass Bounds from Unitarity}
\label{subsec:unitarity}

For particles getting some fixed fraction of their mass from electroweak symmetry breaking, the coupling to the Higgs increases with the mass. Since scattering into Higgses or via Higgs exchange will violate unitarity if this coupling is too large, requiring that the theory be under perturbative control places an upper bound on the mass of the BSM Loryons. To account for these bounds, we impose partial wave unitarity on scattering processes involving Loryons and Higgses. We emphasize that partial wave unitarity does not provide an invariant bound on the Loryon parameter space, and the detailed bound is sensitive to conventions. However, it does provide a useful indication of the region of parameter space in which perturbation theory remains valid. If Loryons exist outside of the region delineated by unitarity, their properties are likely to be poorly described by the treatment presented here. For instance, the formation of bound states is likely to be interesting and merits further study.

Of course, there are other theoretical bounds that may be placed on the couplings of new particles to the Higgs. Chief among these is the vacuum (in)stability of the Higgs: new particles interacting strongly with the Higgs may cause the Higgs self-coupling to run negative at lower scales than in the Standard Model, leading to a prohibitively short lifetime of the metastable vacuum or an outright instability. However, these concerns may be mitigated by additional UV physics (whether in the form of additional particles -- not necessarily Loryons -- or irrelevant operators contributing to the Higgs potential). As such, in what follows we focus on bounds from perturbative unitarity, but considerations involving Higgs vacuum stability would be an interesting target for further exploration.

\begin{figure}[t!]
\centering
\includegraphics[width=0.5\textwidth]{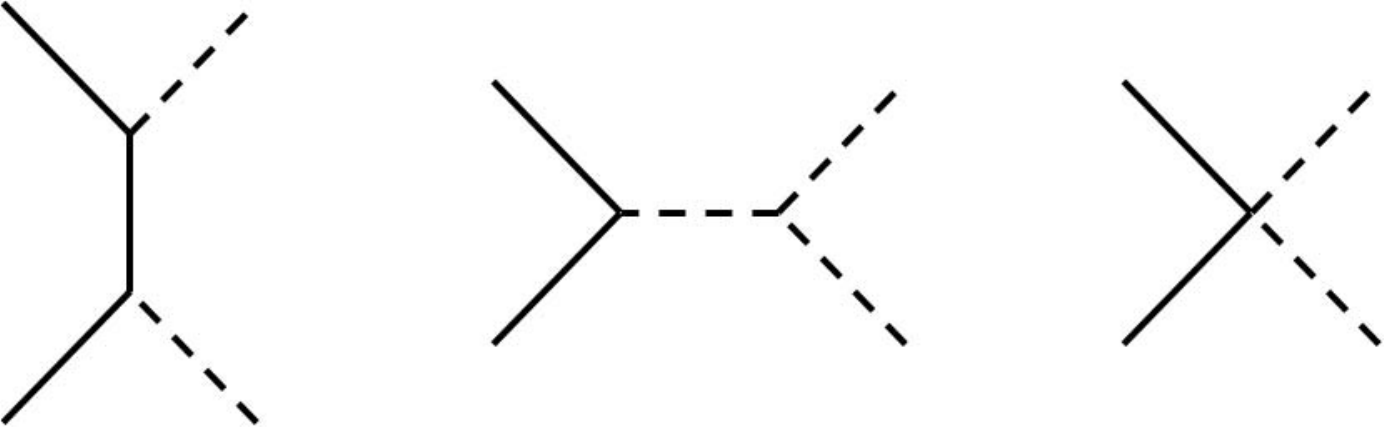}\\[20pt]
\includegraphics[width=0.3\textwidth]{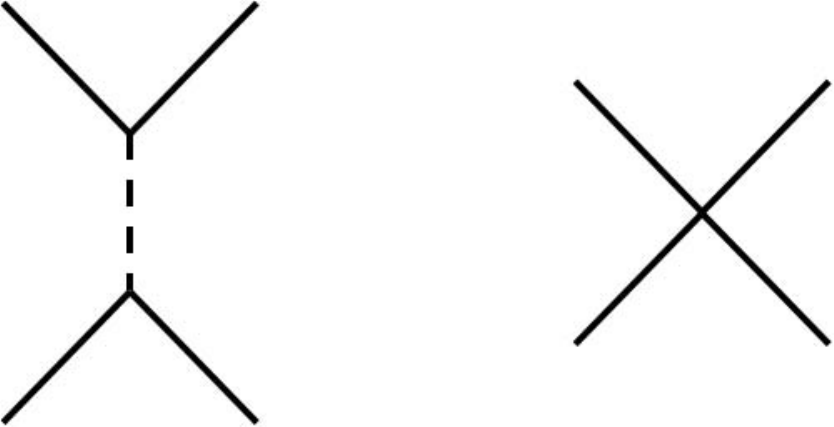}
\caption{Tree-level 2-to-2 scattering processes considered for placing an upper bound on Loryon masses from perturbative unitarity: (upper) Loryon pair scattering to a Higgs pair, and (lower) Loryon pair scattering to a Loryon pair. Dashed lines denote physical Higgs boson $h$ and solid lines are for Loryons. Crossed channels are not explicitly drawn. The last diagram in each row is only available for scalar Loryons assuming renormalizable interactions. For each process, the $t$-channel exchange diagram dominates the contribution.}\label{fig:channels}
\end{figure}

We focus on 2-to-2 scattering processes at the tree level for simplicity. The strongest bounds on the Loryon masses come from the scattering of a Loryon pair into a Higgs pair, or a Loryon pair into a Loryon pair, as depicted in \cref{fig:channels}. For technical simplicity, only contributions from the exchange of scalars and fermions are considered. Diagrams with SM vector boson exchange give subdominant effects.

\begin{figure}[t!]
\centering
\includegraphics[width=0.45\textwidth]{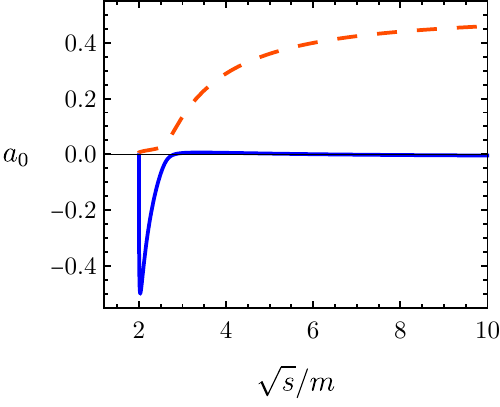}
\caption{An example of the behavior of the zeroth partial wave coefficients $a_0\left(\sqrt{s}\right)$. The plot is for a neutral singlet scalar Loryon $S_0$ getting all of its mass $m=525$ GeV from the Higgs. The two curves correspond to the two eigenvalues of the 2 by 2 scattering matrix $\mqty(S_0 S_0, hh)^T \rightarrow \mqty(S_0 S_0, hh)^T$.}
\label{fig:a0}
\end{figure}

For a general 2-to-2 scattering process taking an initial state $i$ to a final state $f$, one can project onto the spin-0 partial wave component by averaging over the scattering angle with normalization appropriate for arbitrary values of $s$:\footnote{When fermions are involved in the initial or final states, in principle one also needs to project the spinor part onto states with definite helicities using the Wigner $d$-matrix \cite{Jacob:1959at}, although when projecting into an overall spin 0 state the formula \cref{eqn:swave} remains valid for the collision of two spinning particles of equal helicity.}
\begin{align}
a_0 \left(\sqrt{s}\right) = \sqrt{\frac{4\left|\vec p_i\right|\left|\vec p_f\right|}{2^{\d_i+\d_f}s}} \frac{1}{32\p} \int_{-1}^1\dd(\cos\q)\, \mathcal{M}(i\to f) \,,
\label{eqn:swave}
\end{align}
where $\d_{i}/\d_{f}$ is 1 if the initial/final state particles are identical and 0 otherwise. Unitarity of the $S$ matrix then imposes the bound
\begin{align}
\left| \text{Re}(a_0) \right| \le \frac12 \,.
\end{align}
Of particular note is that the bound applies for all values of the center-of-mass energy $\sqrt{s}$, not just in the high energy limit \cite{Goodsell:2018tti}. For example, in the 2-to-2 scattering of heavy Loryons via $t$-channel exchange of a Higgs (bottom-left diagram in \cref{fig:channels}), the maximal value of $\left|\Re\left(a_0\right)\right|$ occurs not in the high energy limit but close to the threshold, as illustrated in \cref{fig:a0}. This channel typically gives the strongest upper bound on the Loryon masses.

For \textbf{scalars}, the limit we derive on its mass depends on the quartic self-coupling of the new field. An upper bound on the quartic self-coupling can be obtained from the high energy limit; we then report the weakest bound after marginalizing over the allowed self-couplings. The final result depends on the representation of the Loryon as well as the fraction of mass that it gets from electroweak symmetry breaking. For a singlet Loryon, the upper bound varies smoothly from 530 to 810 GeV as $f_\text{max}$ goes from 1 to 1/2. For larger representations, the bounds are stronger by up to $\mathcal O(50 \, \mathrm{GeV})$, depending on the precise representation chosen and amount of mass splitting. For \textbf{fermions}, the upper bound on the heaviest mass varies smoothly from about 470 to 780 GeV as $f_\text{max}$ goes from 1 to 1/2 for all possibilities not ruled out by electroweak precision measurements (see \cref{sec:PEWK}).

In addition to constraints on elastic scattering, there are unitarity constraints on the cross-quartic couplings $\lambda_{h\Phi}^{(\prime)}$, or Yukawa squared $y_{12}^2$, arising from the inelastic process $hh \to \text{Loryon} \, \text{Loryon}$, for which $\sum_\text{states} |a_0|^2 \lesssim 1$. They provide a constraint on $\lambda_{h\Phi}^{(\prime)}$ or $y_{12}^2$ that scales as $N^{-\frac12}$ in the number of states $N$. This is inconsequential for the smaller individual viable custodial irreps plotted below, but is to be borne in mind for larger solutions, particularly when considering their effect on Higgs wavefunction normalization (see \cref{sec:future}).

\section{Higgs Coupling Constraints}
\label{sec:higgs}

Having defined the landscape of Loryons, we now turn to consider the constraints from current experimental data. We begin with Higgs coupling measurements, which are particularly impactful given the non-decoupling nature of Loryons. By assumption, our BSM Loryons have an approximate $\mathbb{Z}_2$ symmetry and therefore would only correct Higgs couplings starting from one-loop order. This makes the following measurements potentially important:
\begin{equation}
h\g\g\;\; \text{coupling}\,,\quad
hgg\;\; \text{coupling}\,,\quad
h \rightarrow \text{invisible or untagged width} \,.
\end{equation}
Although Loryons modifying the couplings $h\g\g$ and $hgg$ generically also modify other Higgs couplings, these latter effects are loop-level corrections to tree-level Standard Model couplings and hence not as significant. The constraints from current $hZ\g$ coupling measurement is also typically subdominant compared to that from $h\g\g$, but could potentially play an important role at HL-LHC in future; we briefly discuss this in \cref{sec:future}, where we will also comment on the potential impact of Higgs wavefunction and self coupling corrections.

\subsection{General Formalism}

The leading contribution to the $h\g\g$ ($hgg$) coupling in the SM occurs via loops of charged (colored) particles that have tree-level couplings to the Higgs. As has long been appreciated, any new charged (colored) particles can run in the loop and modify these couplings \cite{Gunion:1989we}. These modifications can be captured by the parameter $\k_\g$ ($\k_g$), which simply rescales the $h\g\g$ ($hgg$) vertex, with $\k_\g=\k_g=1$ for the SM.

For a scalar Loryon $\phi$ or a fermionic Loryon $\psi$ discussed in \cref{sec:candidates}, we can read off the tree-level Higgs-Loryon-Loryon coupling from \cref{eqn:ScalarLag,eqn:FermionLag}, facilitated by our definitions in \cref{eqn:fScalar,eqn:fFermion}:
\begin{equation}
\lag \supset - \frac{1}{2^\rho}\, f_\phi \, \frac{2 m_\phi^2}{v}\, h\, \phi^\dagger\, \phi - f_\psi\, \frac{m_\psi}{v}\, h\, \overline{\psi}\, \psi \, ,
\end{equation}
where $f_i$ is $f_V$ for the $i$th particle. Generally, particles coupled to the Higgs in this form contribute to $\k_\g$ and $\k_g$ as \cite{Carmi:2012in}
\begin{subequations}\label{eqn:HiggsDecaysLoop}
\begin{align}
\k_\g &\propto \sum_i f_i\, Q_i^2\, A_{s_i}(\t_i) \,, \\[8pt]
\k_g  &\propto \sum_i f_i\, C_i\, A_{s_i}(\t_i) \,.
\end{align}
\end{subequations}
The sum runs over all contributing complex scalars, Dirac fermions, or vector bosons. For each contributing particle $i$, $Q_i$ denotes its electromagnetic charge; $C_i$ is the Dynkin index of its $SU(3)_C$ representation, namely $\tr\left(t_i^A t_i^B\right) = C_i \delta^{AB}$ with $t_i^A$ denoting the $SU(3)_C$ generators; $\t_i=4m_i^2/m_h^2$ parameterizes the mass; $s_i$ denotes the spin; and the spin-dependent form factors $A_{s_i}(\t)$ are given by
\begin{subequations}\label{eqn:formfactors}
\begin{align}
A_0(\t) &= \t \big[ 1 - \t F(\t) \big] \,, \\[6pt]
A_{1/2}(\t) &= -2\t \big[ 1 + (1-\t) F(\t) \big] \,, \\[6pt]
A_1(\t) &= 2+3\t \big[ 1 + (2-\t) F(\t) \big] \,,
\end{align}
\end{subequations}
with
\begin{equation}
F(\t) = \left\{\begin{array}{ll}
\arcsin^2\left( 1/\sqrt{\t} \right) &\qquad \t\ge 1 \\[5pt]
-\frac14 \left[ \log\frac{1+\sqrt{1-\t}}{1-\sqrt{1-\t}} - i\pi \right]^2 &\qquad \t<1
\end{array}\right. \,.
\label{eqn:Ftau}
\end{equation}
When the particles running in the loop are asymptotically heavy, namely $\t_i\to\infty$, these form factors asymptote to constants:
\begin{align}
A_0 \to -1/3 \,,\qquad   A_{1/2} \to -4/3 \,,\qquad   A_1 \to 7 \,.
\label{eqn:asymptote}
\end{align}
In practice, this is a good approximation ($\lesssim 10\%$ error) for particles heavier than the Higgs, which is the case for most BSM Loryons of interest.

\subsection{$h\g\g$ Coupling}
\label{subsec:hgammagamma}

To compute $\k_\g$ for a particular BSM model, we can simply use (see \cref{eqn:HiggsDecaysLoop})
\begin{align}
\k_\g = 1 + \frac{\sum_\text{BSM} f_i\, Q_i^2\, A_{s_i}(\t_i)}{\sum_\text{SM} f_i\, Q_i^2\, A_{s_i}(\t_i)} \,.
\label{eqn:kgamma}
\end{align}
In the SM sum, we include the $W^\pm$ bosons, the top, bottom, charm quarks, and the tau lepton. Contributions from other charged SM particles are negligible due to their tiny form factors.

\begin{figure}[t!]
\centering
\includegraphics[align=c,width=.55\textwidth]{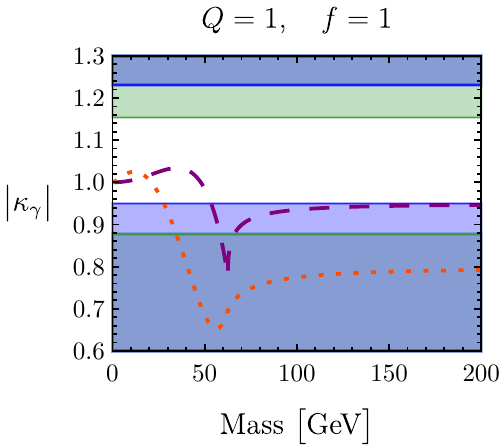}\hspace{20pt}
\includegraphics[align=c,width=.27\textwidth]{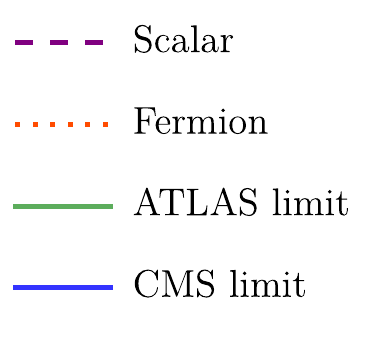}
\caption{The contribution to $\left|\k_\g\right|$ from the addition of a new charge-1 complex scalar or Dirac fermion getting all of its mass from electroweak symmetry breaking as a function of the mass of the new particle. Limits on $\left|\k_\g\right|$ from the ATLAS \cite{Aad:2019mbh} and CMS \cite{Sirunyan:2018koj} collaborations are shaded.}
\label{fig:kgammatypical}
\end{figure}

For the experimental constraints on $\k_\g$, we use the most recent ATLAS and CMS measurements. In particular, we use each collaboration's joint fit to $\k_\g$, which holds in the absence of deviations to tree-level Higgs couplings and untagged/invisible Higgs decay width. Upon neglecting these small effects, the $2\si$ allowed region from ATLAS is $\left| \k_\g \right| \in (0.877, 1.15)$ \cite{Aad:2019mbh}, while CMS finds $\left| \k_\g \right| \in (0.949, 1.23)$~\cite{Sirunyan:2018koj}. In \cref{fig:kgammatypical}, we show these bounds against contributions from a typical scalar or fermion Loryon. Note in particular that for a BSM Loryon heavier than the Higgs, the relevant asymptote in \cref{eqn:asymptote} is already effective. Nevertheless, a larger deviation of $\left|\k_\g\right|$ happens near the threshold mass $m_i=m_h/2$.

\begin{figure}[t!]
\centering
\includegraphics[align=c,width=.55\textwidth]{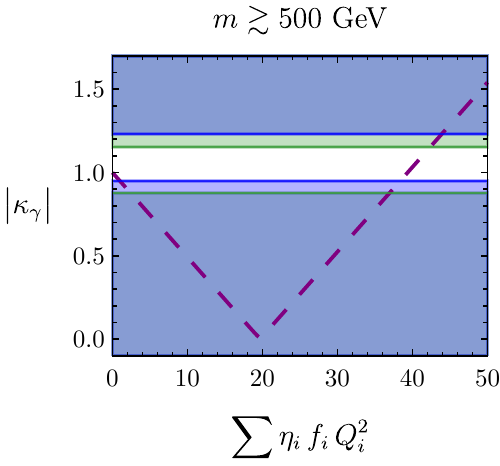}\hspace{20pt}
\includegraphics[align=c,width=.27\textwidth]{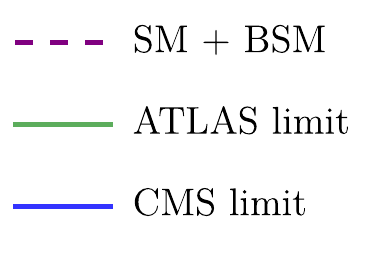}\\
\caption{The contribution to $|\k_\g|$ from new particles heavier than $\sim 500$ GeV as a function of their fraction $f_i$ of mass from EWSB and their electric charge $Q_i$; $\h_i=1\,(4)$ for complex scalars (fermions). Large values of $\sum_i \h_i f_i Q_i^2$ are sufficient to flip the sign of $\k_\g$ while maintaining the same magnitude. ATLAS and CMS limits are shaded.}
\label{fig:kgammaSignFlip}
\end{figure}

Current bounds on $\left|\k_\g\right|$ constrain the sum of the contributions from BSM Loryons
\begin{equation}
\sum_\text{BSM} f_i\, Q_i^2\, A_{s_i}(\t_i)   \;\;\rightarrow\;\;   - \frac13 \sum_\text{BSM} \h_i\, f_i\, Q_i^2 \,,
\label{eqn:kgammaBSMsum}
\end{equation}
with $\h_i=1\,(4)$ for scalars (fermions). One way to satisfy the experimental constraints is of course to ensure that the contribution in \cref{eqn:kgammaBSMsum} sufficiently small. However, there is a second way to satisfy the constraint. Note that the SM contribution to $\k_\g$ is dominated by $W^\pm$ bosons; contributions from BSM scalar and fermionic Loryons (if heavier than half the Higgs mass) would come with an opposite sign compared to this dominant piece in the SM sum. Therefore, there is also a viable window where the magnitude of the second term in \cref{eqn:kgamma} becomes big enough to flip the sign of (the real part of) $\k_\g$, while keeping the magnitude close to the SM-only result. This is illustrated in \cref{fig:kgammaSignFlip}.

To summarize, requiring BSM Loryons to satisfy both the ATLAS and CMS bounds on the $h\g\g$ coupling measurements, we find the constraints
\begin{align}
\sum_i \h_i\, f_i\, Q_i^2 < 0.995 \qquad\text{or}\qquad
\sum_i \h_i\, f_i\, Q_i^2 \in (38.4, 42.4) \,,
\label{eq:gammalims}
\end{align}
for asymptotically heavy BSM Loryons, with $\h_i=1\,(4)$ for complex scalars (fermions). If the BSM Loryons are not asymptotically heavy, they will contribute more. Therefore, for lighter Loryons, the first limit would become stronger while the second would shift to a window with lower values (see \cref{fig:scalarcombo}).

\begin{table}[t!]
\renewcommand{\arraystretch}{1.6}
\setlength{\tabcolsep}{0.5em}
\setlength{\arrayrulewidth}{1.2pt}
\centering
\begin{tabular}{ c | c | c | c | c | c | c  }
Field & $[1,1]_1$ & $[2,2]_0$ & $[3,3]_0$ & $[2,3]_{-1/2}$ & $[2,1]_{1/2}\oplus[1,2]_{1/2}$ & $[1,3]_0\oplus[2,2]_0$ \\
\hline
$\sum\h_i\, Q_i^2$ & $1$ & $1$ & $6$ & $7$ & $8$ & $16$
\end{tabular}
\caption{Values of $\sum \h_i\, Q_i^2$ for some possible BSM Loryons with $f_\text{max}=1$. The first four entries are scalars; the last two are fermions.}
\label{tab:etaQ2}
\end{table}

\begin{table}[t]
\renewcommand{\arraystretch}{1.6}
\setlength{\tabcolsep}{0.7em}
\setlength{\arrayrulewidth}{1.2pt}
\centering
\begin{tabular}{  c | c | c | c | c | c | c | c | c  }
\multicolumn{9}{c}{\textsc{Scalar Scorecard}}\\[5pt]
 & $R=1$ & 2 & 3 & 4 & 5 & 6 & 7 & 8 \\
\hline
$L=1$ & $|Y_{max}|=1,\sim\!3$ & $\frac12,\sim\!\frac52$ & $0,\sim\!2$ & $\sim\!\frac32$ & $\sim\!1$ & $\sim\!\frac12$ & $\sim\!0$ & $\times$ \\
\hline
2 & $\frac12,\sim\!\frac72$ & $1,\sim\!4$ & $\frac32,\sim\!\frac72$ & $1,\sim\!3$ & $\frac12,\sim\!\frac32$ & $1$ & $\frac12$ & $0$ \\
\hline
3 & $0,\sim\!3$ & $\frac32,\sim\!\frac72$ & $1,\sim\!2$ & $\frac12,\sim\!\frac32$ & $0,\sim\!1$ & $\frac12$ & $0$ & $\times$ \\
\hline
4 & $\sim\!\frac72$ & $1,\sim\!3$ & $\frac12,\sim\!\frac32$ & $\sim\!1$ & $\sim\!\frac12$ & $\sim\!0$ & $\times$ & $\times$ \\
\hline
5 & $\sim\!3$ & $\frac12,\sim\!\frac32$ & $0,\sim\!1$ & $\sim\!\frac12$ & $\sim\!0$ & $\times$ & $\times$ & $\times$ \\
\hline
6 & $\sim\!\frac52$ & $1$ & $\frac12$ & $\sim\!0$ & $\times$ & $\times$ & $\times$ & $\times$ \\
\hline
7 & $\sim\!2$ & $\frac12$ & $0$ & $\times$ & $\times$ & $\times$ & $\times$ & $\times$ \\
\hline
8 & $\sim\!\frac32$ & $0$ & $\times$ & $\times$ & $\times$ & $\times$ & $\times$ & $\times$
\end{tabular}\\[13pt]
\caption{The representations of scalar BSM Loryons still viable after considering constraints on $\k_\g$. A $\sim$ means the representation requires flipping the sign of $\k_\g$.}
\label{tab:scalarScorekgamma}
\end{table}
\begin{table}[t!]
\renewcommand{\arraystretch}{1.6}
\setlength{\tabcolsep}{0.7em}
\setlength{\arrayrulewidth}{1.2pt}
\centering
\begin{tabular}{  c | c | c | c | c | c | c | c | c  }
\multicolumn{9}{c}{\textsc{Vector-like Fermion Scorecard}}\\[5pt]
 & $R=1$ & 2 & 3 & 4 & 5 & 6 & 7 & 8 \\
\hline
$L=1$ & $|Y_{max}|=0,\sim\!2$ & $\sim\!\frac32$ & $\sim\!1$ & $\times$ & $\times$ & $\times$ & $\times$ & $\times$ \\
\hline
2 & $\sim\!\frac32$ & $0,\sim\!2$ & $\sim\!\frac32$ & $\sim\!1$ & $\times$ & $\times$ & $\times$ & $\times$ \\
\hline
3 & $\sim\!1$ & $\sim\!\frac32$ & $\sim\!0$ & $\times$ & $\times$ & $\times$ & $\times$ & $\times$ \\
\hline
4 & $\times$ & $\sim\!1$ & $\times$ & $\times$ & $\times$ & $\times$ & $\times$ & $\times$ \\
\hline
5 & $\times$ & $\times$ & $\times$ & $\times$ & $\times$ & $\times$ & $\times$ & $\times$ \\
\hline
6 & $\times$ & $\times$ & $\times$ & $\times$ & $\times$ & $\times$ & $\times$ & $\times$ \\
\hline
7 & $\times$ & $\times$ & $\times$ & $\times$ & $\times$ & $\times$ & $\times$ & $\times$ \\
\hline
8 & $\times$ & $\times$ & $\times$ & $\times$ & $\times$ & $\times$ & $\times$ & $\times$
\end{tabular}\\[13pt]
\caption{The representations and corresponding fields for the vector-like fermionic BSM Loryons still viable after considering constraints on $\k_\g$. A $\sim$ means the representation requires flipping the sign of $\k_\g$.}
\label{tab:fermionScorekgamma}
\end{table}

Values of $\sum_i \h_i \, Q_i^2$ for select Loryons are listed in \cref{tab:etaQ2}. The impact of the constraints in \cref{eq:gammalims} on scalar and fermionic Loryons in various custodial irreps is indicated by the ``scorecards'' of \cref{tab:scalarScorekgamma} and \cref{tab:fermionScorekgamma}, respectively. (In what follows, these scorecards will be updated as successive constraints are taken into account.) If the new Loryons receive all or almost all of their mass from their coupling to the Higgs, the first limit in  \cref{eq:gammalims} can be satisfied by adding at most one charge-1 scalar (and possibly some neutral particles). It can also be satisfied by adding a scalar representation with a large mass splitting. Since the largest representation of the unbroken $SU(2)_V$ receives a mass shift opposite in sign to the others, a large mass splitting will drive the value of $f_V$ towards 0 for the particles in the largest representation, thereby satisfying the limit on $\k_\g$. Other possibilities, including larger electroweak representations, representations with larger hypercharge, and fermionic representations, make a large enough contribution to \cref{eqn:kgammaBSMsum} that flipping the sign of $\k_\g$ is required. Among other things, this implies that the only way for fermionic Loryons to satisfy constraints on $\k_\g$ is to flip the sign, as indicated in \cref{tab:fermionScorekgamma}.

\subsection{$hgg$ Coupling}
\label{subsec:hgg}

\begin{table}[h!]
\renewcommand{\arraystretch}{1.6}
\setlength{\tabcolsep}{0.7em}
\setlength{\arrayrulewidth}{1.2pt}
\centering
\begin{tabular}{ c | c | c | c }
Field & $\omega_{|3Y|}$ & $\Pi_{|6Y|}$ & $P_{|3Y|}$ \\
\hline
$\sum \h_i\, C_i$ & $1/2$ & $1$ & $2$ \\
\hline
$\sum \h_i\, Q_i^2$ & $3Y^2$ & $6Y^2+3/2$ & $12Y^2$
\end{tabular}
\caption{Values of $\sum \h_i\, Q_i^2$ and $\sum \h_i\, C_i$ for some possible new SM representations.}
\label{tab:kgammakg}
\end{table}

\begin{figure}[t!]
\centering
\includegraphics[width=0.6\textwidth]{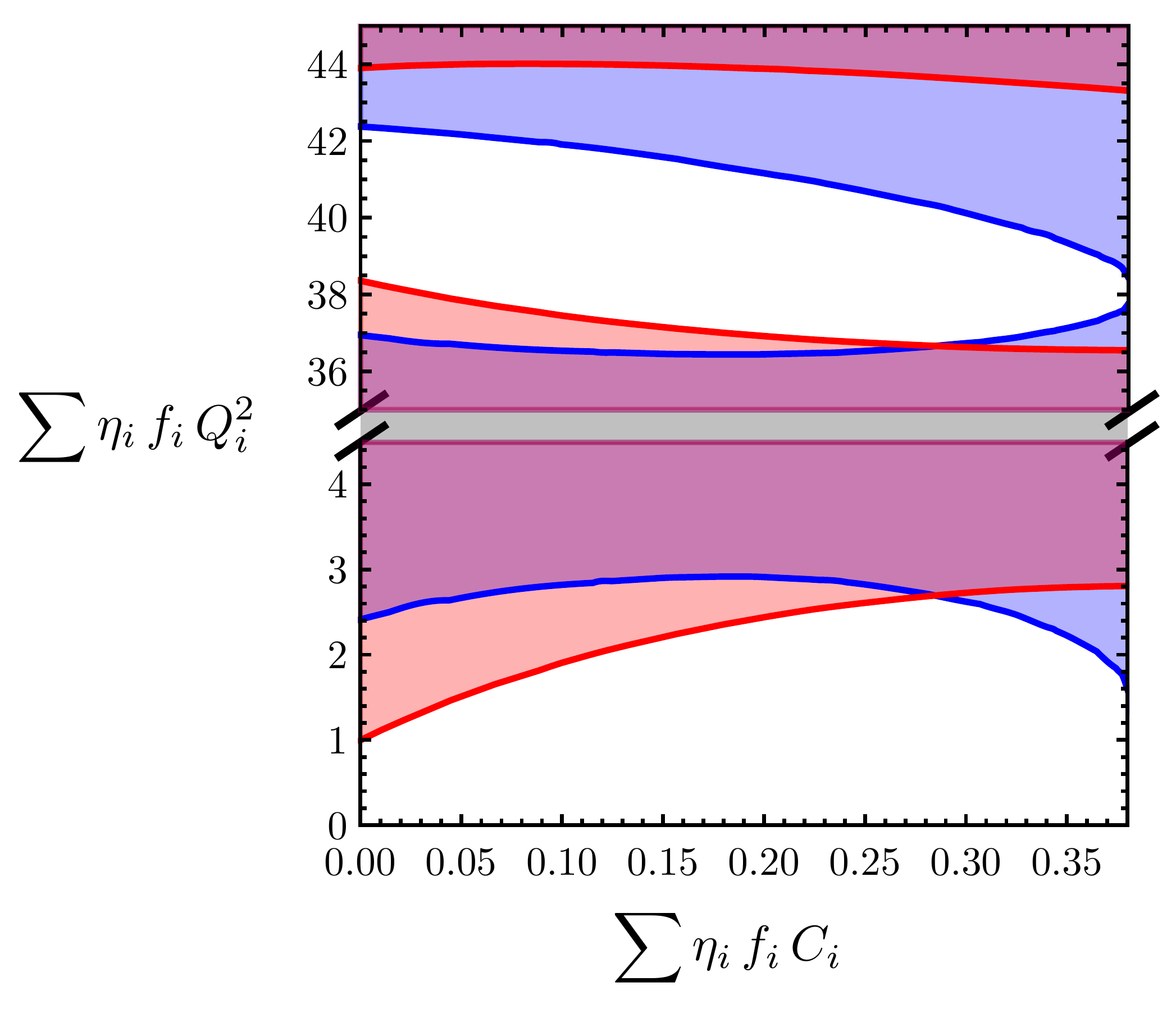}\hspace{55pt}
\caption{The allowed electric charge $Q_i$ as a function of the color charges $C_i$ of new fields due to $\k_\g-\k_g$ constraints. ATLAS constraints \cite{Aad:2019mbh} are shaded blue, CMS constraints \cite{Sirunyan:2018koj} are shaded red. }
\label{fig:kgammakg}
\end{figure}

As with $\k_\g$, we can compute $\k_g$ for a particular BSM model simply using (see \cref{eqn:HiggsDecaysLoop})
\begin{equation}
\kappa_g = 1 + \frac{\sum_\text{BSM} f_i\, C_i\, A_{s_i}(\t_i)}{\sum_\text{SM} f_i\, C_i\, A_{s_i}(\t_i)} \,.
\end{equation}
In the SM sum, we include the top, bottom, and charm quarks; contributions from other quarks are negligible due to their tiny form factors. For $\k_g$, BSM scalar and fermionic Loryons (if heavier than half the Higgs mass) contribute with the same sign as the dominant piece (from the top quark) in the SM sum, so there is no flipping sign option and the sum of the BSM Loryon contributions must not be too large. We constrain these contributions at the $2\si$ level using the results from ATLAS and CMS expressed in the $\k_\g$ vs $\k_g$ plane \cite{Aad:2019mbh, Sirunyan:2018koj}, again neglecting deviations to tree-level Higgs couplings and new untagged or invisible Higgs decay width. The bounds are translated into the constraints on the $\sum \h_i f_i Q_i^2$ vs $\sum \h_i f_i C_i$ plane in \cref{fig:kgammakg}. We find that these constraints essentially exclude all colored Loryons, except for scalar Loryons of the SM representations $(3,1)_Y$ with $|Y|\le 1$, as summarized in \cref{tab:kgammakg}.

\subsection{$h \rightarrow$ invisible or untagged}

Loryons lighter than half the Higgs mass will also make new channels for the Higgs to directly decay at the tree level. For each scalar Loryon $\phi$ or Dirac fermion Loryon $\psi$, the partial width is
\begin{subequations}\label{eqn:HiggsDecaysTree}
\begin{align}
\Gamma_{h\to\phi\phi^{(\dagger)}} &= f_\phi^2\, \frac{G_F\, m_h\, m_\phi^2}{2^\rho \cdot 8\sqrt{2}\,\pi}\, \frac{4\,m_\phi^2}{m_h^2}\, \left( 1 - \frac{4\,m_\phi^2}{m_h^2} \right)^{1/2} \,, \\[8pt]
\Gamma_{h\to\psi\bar\psi} &= f_\psi^2\, \frac{G_F\, m_h\, m_\psi^2}{4\sqrt{2}\,\pi} \left( 1 - \frac{4\, m_\psi^2}{m_h^2} \right)^{3/2} \,.
\end{align}
\end{subequations}

\begin{figure}[t!]
\centering
\includegraphics[align=c,width=0.45\textwidth]{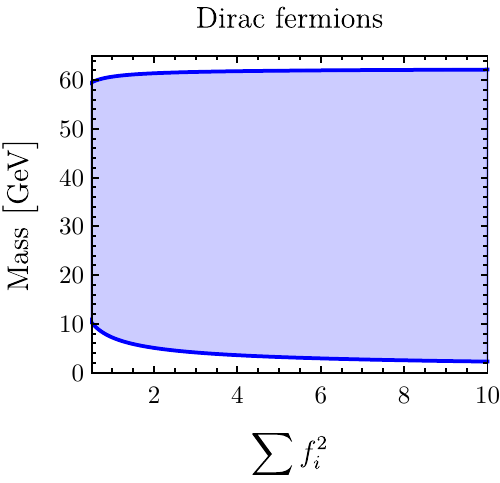}\hspace{40pt}
\includegraphics[align=c,width=0.45\textwidth]{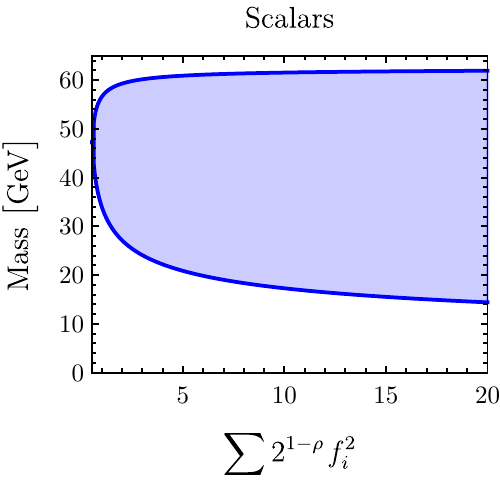}
\caption{Allowed region (unshaded) in the mass vs $\sum f_i^2$ $\big(\sum 2^{1-\rho}\s f_i^2\big)$ plane for new Dirac fermion (scalar) particles whose mass is less than half the Higgs mass, arising from the upper limit on the Higgs branching ratio to `untagged'. The plot assumes all new particles have the same mass.}
\label{fig:untagged}
\end{figure}

Constraints on new Higgs decay channels depend on the properties and fate of the new particles. If the particle is neutral and detector stable, or if it decays promptly into neutral and detector-stable final states, it would contribute to the Higgs invisible decay width. If it decays promptly into visible final states, it generally contributes to the Higgs ``untagged'' decay width, which is less constrained than the invisible width (though sufficiently distinctive final states can lead to stronger constraints). There are more exotic possibilities, such as long-lived decays, but these are typically more strongly constrained than invisible or untagged decays. For the purposes of the current discussion, we assume that Loryons which are lighter than half the Higgs mass contribute to the ``untagged'' decay width of the Higgs. In this case, the branching ratio to the new decay channels must be less than 0.47~\cite{Sirunyan:2018koj}, which constrains the size of the BSM Loryon contribution to the Higgs partial widths in \cref{eqn:HiggsDecaysTree}. For fixed values of $f_i^2$, the mass of the Loryon has to be either small enough or close enough to the kinematic threshold, such that the partial width is not too big. Allowed regions for scalar and fermionic Loryon are shown in \cref{fig:untagged}.

\section{Precision Electroweak Constraints}
\label{sec:PEWK}

Extending the SM with new particles that carry electroweak quantum numbers could potentially be subject to strong constraints from electroweak precision measurements. In the case of the BSM Loryon models studied here, the oblique framework \cite{Lynn:1985fg,Peskin:1991sw} is a good approximation since they interact with SM primarily through the Higgs and gauge bosons. The leading order (one-loop) corrections therefore modify the electroweak gauge boson self-energies, which up to $\mathcal{O}(p^4)$ can be parameterized by the seven (extended) electroweak parameters $S, T, U, V, W, X, Y$ \cite{Maksymyk:1993zm,Burgess:1993mg,Kundu:1996ah,Barbieri:2004qk}. Among these seven parameters, $S, T, W, Y$ arguably capture the most important effects as they are the leading ones in their respective symmetry classes \cite{Barbieri:2004qk}. Since we are considering custodially symmetric BSM Loryon models, the correction to $T$ (and also its higher $\mathcal{O}(p^2)$ analog $U$) vanishes at one loop order. This leaves us to focus on the parameters $S$, $W$, and $Y$:
\begin{subequations}
\begin{align}
S &= - \frac{4\cos\theta_W \sin\theta_W}{\alpha}\, \Pi'_{3B} (p^2=0) \,, \\[5pt]
W &= - \frac12\, m_W^2\, \Pi''_{33} (p^2=0) \,, \\[5pt]
Y &= - \frac12\, m_W^2\, \Pi''_{BB} (p^2=0) \,,
\end{align}
\end{subequations}
where the $\Pi$s are the gauge boson self energies, primes denoting differentiation with respect to $p^2$.

For a scalar Loryon of the custodial irrep $[L, R]_Y$, we find its contributions given by
\begin{subequations}\label{eqn:SWYScalar}
\begin{align}
\Delta S &= \frac{2}{\pi}\, \sum_{i,j=1}^n T_{ij}^3\, Y_{ji}\, \Pi'_S \left( m_i, m_j \right) \,, \\[5pt]
\Delta W &= m_W^2\, \frac{g_2^2}{16\pi^2}\, \sum_{i,j=1}^n T_{ij}^3\, T_{ji}^3\, \Pi''_S \left( m_i, m_j \right) \,, \\[5pt]
\Delta Y &= m_W^2\, \frac{g_1^2}{16\pi^2}\, \sum_{i,j=1}^n Y_{ij}\, Y_{ji}\, \Pi''_S \left( m_i, m_j \right) \,,
\end{align}
\end{subequations}
where $n=LR$ denotes the total number of components of the Loryon, and the scalar form factors are
\begin{subequations}
\begin{align}
\Pi'_S \left( m_i, m_j \right) &= \frac{1}{2^{\rho_i}}\,\int_0^1 \dd x\, \Bigg[ x(1-x) \log \frac{\mu^2}{x m_i^2 + (1-x) m_j^2} \Bigg] \,, \label{eqn:PiSp} \\[5pt]
\Pi''_S \left( m_i, m_j \right) &= \frac{1}{2^{\rho_i}}\, \int_0^1 \dd x\, \frac{x^2(1-x)^2}{x m_i^2 + (1-x) m_j^2} \,. \label{eqn:PiSpp}
\end{align}
\end{subequations}
Note that $T_{ij}^3$ and $Y_{ij}$ are (elements of) the SM generators in the mass basis of the Loryon representation, which are generically not diagonal due to the mass mixings among different gauge eigenstates.

We see that $\Delta S$ is only nonzero due to the mass splitting among the Loryon components, namely that $\Pi'_S \left( m_i, m_j \right)$ is not the same for all $i,j$; terms that are independent of $i,j$ (such as the RG scale $\mu$ dependence) will drop upon the sum, as they yield $\tr\left(T^3\, Y\right)=0$. On the other hand, $\Delta W$ and $\Delta Y$ do not depend on the mass splitting, and can in principle constrain any custodial irrep of Loryons. However, their values are typically more than one order of magnitude smaller than $\Delta S$, due to a combination of the extra mass suppression factor $m_W^2/m_i^2$, and the smallness of the form factor in \cref{eqn:PiSpp}. This makes the current constraints from $W$ and $Y$ parameters \cite{Farina:2016rws} numerically unimportant for the Loryon mass range of our interest.

Moving on to the fermionic cases, we focus on the $S$ parameter. We find that a pair of fermionic Loryons of the custodial irreps $[L_1, R_1]_Y$ and $[L_2, R_2]_Y$ contribute to $S$ as
\begin{equation}
\Delta S = \frac{4}{\pi}\, \sum_{i,j=1}^n T_{ij}^3\, Y_{ji} \bigg[\, \xi_\Sigma\, \Pi'_{F,\,\Sigma} \left( m_i, m_j \right) + \xi_\Delta\, \Pi'_{F,\,\Delta} \left( m_i, m_j \right) \bigg] \,,
\label{eqn:SFermion}
\end{equation}
with $n=L_1R_1+L_2R_2$ denoting the total number of Dirac fermions and the form factors
\begin{subequations}
\begin{align}
\Pi'_{F,\,\Sigma} \left( m_i, m_j \right) &= \int_0^1 \dd x\, \Bigg[ 2\log \frac{\mu^2}{x m_i^2 + (1-x) m_j^2} - 1 \Bigg] x(1-x) \,, \\[5pt]
\Pi'_{F,\,\Delta} \left( m_i, m_j \right) &= \int_0^1 \dd x\, \frac{m_i m_j}{x m_i^2 + (1-x) m_j^2}\, x(1-x) \,.
\end{align}
\end{subequations}
Here our notation in \cref{eqn:SFermion} allows for a generic coupling between a Dirac fermion $\psi$ and a gauge boson $V$:
\begin{equation}
\lag \supset g\, \bar\psi\, \gamma^\mu \left( \xi_V - \xi_A \gamma^5 \right) t^a\, \psi\, V_\mu^a \,,
\end{equation}
and $\xi_\Sigma$ ($\xi_\Delta$) tracks the contributions from vertex insertions with the same (opposite) chiralities:
\begin{equation}
\xi_\Sigma = \xi_{1,V} \xi_{2,V} + \xi_{1,A} \xi_{2,A} \,,\qquad
\xi_\Delta = \xi_{1,V} \xi_{2,V} - \xi_{1,A} \xi_{2,A} \,.
\end{equation}
Applying it to our case of vector-like fermions, we can aggregate the contributions from all the chiral component insertions which all share the same form of the generator; this effectively leads us to plugging in $\xi_\Sigma = \xi_\Delta = 1$.

\begin{table}[t]
\renewcommand{\arraystretch}{1.6}
\setlength{\tabcolsep}{0.7em}
\setlength{\arrayrulewidth}{1.2pt}
\centering
\begin{tabular}{ c | c | c | c | c | c }
Rep & $[2,2]_0$ & $[3,3]_0$ & $[4,4]_0$ & $[2,4]_0$ & $[2,3]_{-1/2}$ \\
\hline
Allowed $r_\text{split}$ & $(-.67,1.98)$ & $(-.25,.28)$ & $(-.05,.05)$ & $(-.36,.44)$ & $(-.46,.58)$
\end{tabular}
\caption{Examples of the magnitude of allowed mass-splitting \cref{eqn:rsplit} if a particular scalar representation is the only new field contributing to $S$ parameter. \label{tab:scalarsparam}}
\end{table}

\begin{table}[t]
\renewcommand{\arraystretch}{1.6}
\setlength{\tabcolsep}{0.3em}
\setlength{\arrayrulewidth}{1.2pt}
\centering
\begin{adjustbox}{tabular=c | c | c | c | c | c | c  ,center}
\text{Rep} & $[2,1],[1,2]_{1/2}$ & $[2,2],[1,1]_0$ & $[3,1],[2,2]_0$ & $[1,3],[2,2]_0$ & $[2,3],[1,2]_{1/2}$ & $[3,2],[2,3]_{1/2}$ \\
\hline
Allowed $f_\text{max}$ & $(.66,1)$ & $\emptyset$ & $(.71,1)$ & $(.71,1)$ & $(.51,.58)$ & $(.68,.81)$\\
\hline
$\sum\h_if_iQ_i^2$ & $(11,8)$ & $\emptyset$ & $(19,16)$ & $(19,16)$ & $(207,17)$ & $(109,61)$
\end{adjustbox}
\caption{Examples of the allowed $f_\text{max}$ for new fermionic fields if they are the only new contribution to the $S$ parameter. Also shown is the range of values of $\sum\h_if_iQ_i^2$ for the given range of $f_\text{max}$. The vector-like mass is taken to be the same for both fermions involved in the Yukawa interaction; this gives a weaker bound than allowing two different vector-like masses.}
\label{tab:fermionsparam}
\end{table}

We take the current $2\s \sigma$ bound on $S$ to be 0.14 \cite{ParticleDataGroup:2018ovx}. Note that this corresponds to the projection of the combined $S,T$ fit onto the $S$ axis, rather than the $2\s \sigma$ bound on $S$ with $T=0$, which would lead to a tighter bound. Although we focus on custodial multiplets to minimize contributions to $T$, small amounts of soft custodial symmetry breaking in the explicit mass terms allow most of the positive region of the $S,T$ ellipse to be explored. For each choice of representation, we consider the limits placed by requiring that the contribution to $S$ from the Loryons obey this bound.\footnote{We emphasize that it is possible to go beyond our minimal models to include multiple Loryons that are in various representations.  In particular, a representation can yield a negative contribution to $S$, and so it is possible to exceed the limits in \cref{tab:scalarsparam} and \cref{tab:fermionsparam} with judicious choice of representations.}

\begin{table}[t]
\renewcommand{\arraystretch}{1.6}
\setlength{\tabcolsep}{0.7em}
\setlength{\arrayrulewidth}{1.2pt}
\centering
\begin{tabular}{  c | c | c | c | c | c | c | c | c  }
\multicolumn{9}{c}{\textsc{Scalar Scorecard}}\\[5pt]
 & $R=1$ & 2 & 3 & 4 & 5 & 6 & 7 & 8 \\
\hline
$L=1$ & $|Y_{max}|=1,\sim\!3$ & $\frac12,\sim\!\frac52$ & $0,\sim\!2$ & $\sim\!\frac32$ & $\sim\!1$ & $\sim\!\frac12$ & $\sim\!0$ & $\times$ \\
\hline
2 & $\frac12,\sim\!\frac72$ & $1,\sim\!4$ & $\frac12,\sim\!\frac72$ & $0,\sim\!3$ & $\sim\!\frac32$ & $\sim\!1$ & $\sim\!\frac12$ & $\sim\!0$ \\
\hline
3 & $0,\sim\!3$ & $\frac12,\sim\!\frac72$ & $0,\sim\!2$ & $\sim\!\frac32$ & $\sim\!1$ & $\sim\!\frac12$ & $\sim\!0$ & $\times$ \\
\hline
4 & $\sim\!\frac72$ & $0,\sim\!3$ & $\sim\!\frac32$ & $\sim\!1$ & $\sim\!\frac12$ & $\sim\!0$ & $\times$ & $\times$ \\
\hline
5 & $\sim\!3$ & $\sim\!\frac32$ & $\sim\!1$ & $\sim\!\frac12$ & $\sim\!0$ & $\times$ & $\times$ & $\times$ \\
\hline
6 & $\sim\!\frac52$ & $\sim\!1$ & $\sim\!\frac12$ & $\sim\!0$ & $\times$ & $\times$ & $\times$ & $\times$ \\
\hline
7 & $\sim\!2$ & $\sim\!\frac12$ & $\sim\!0$ & $\times$ & $\times$ & $\times$ & $\times$ & $\times$ \\
\hline
8 & $\sim\!\frac32$ & $\sim\!0$ & $\times$ & $\times$ & $\times$ & $\times$ & $\times$ & $\times$
\end{tabular}\\[13pt]
\caption{The representations of scalar BSM Loryons still viable after considering constraints on $S$. A $\sim$ means the representation is not viable on its own but can be added together with other BSM representations to flip the sign of $\kappa_\gamma$.}
\label{tab:scalarScoreS}
\end{table}

\begin{table}[t!]
\renewcommand{\arraystretch}{1.6}
\setlength{\tabcolsep}{0.7em}
\setlength{\arrayrulewidth}{1.2pt}
\centering
\begin{tabular}{  c | c | c | c | c | c | c | c | c  }
\multicolumn{9}{c}{\textsc{Vector-like Fermion Scorecard}}\\[5pt]
 & $R=1$ & 2 & 3 & 4 & 5 & 6 & 7 & 8 \\
\hline
$L=1$ & $|Y_{max}|=\times$ & $\sim\!\frac32$ & $\sim\!1$ & $\times$ & $\times$ & $\times$ & $\times$ & $\times$ \\
\hline
2 & $\sim\!\frac32$ & $\sim\!1$ & $\sim\!\frac12$ & $\times$ & $\times$ & $\times$ & $\times$ & $\times$ \\
\hline
3 & $\sim\!1$ & $\sim\!\frac12$ & $\times$ & $\times$ & $\times$ & $\times$ & $\times$ & $\times$ \\
\hline
4 & $\times$ & $\times$ & $\times$ & $\times$ & $\times$ & $\times$ & $\times$ & $\times$ \\
\hline
5 & $\times$ & $\times$ & $\times$ & $\times$ & $\times$ & $\times$ & $\times$ & $\times$ \\
\hline
6 & $\times$ & $\times$ & $\times$ & $\times$ & $\times$ & $\times$ & $\times$ & $\times$ \\
\hline
7 & $\times$ & $\times$ & $\times$ & $\times$ & $\times$ & $\times$ & $\times$ & $\times$ \\
\hline
8 & $\times$ & $\times$ & $\times$ & $\times$ & $\times$ & $\times$ & $\times$ & $\times$
\end{tabular}\\[13pt]
\caption{The representations and corresponding fields for the vector-like fermion BSM Loryons still viable after considering constraints on $S$. A $\sim$ means the representation is not viable on its own but can be added together with other BSM representations to flip the sign of $\kappa_\gamma$.}
\label{tab:fermionScoreS}
\end{table}

For \textbf{scalars}, the contribution to $S$ is only non-zero if there is a mass splitting among the Loryon states.  We therefore report the magnitudes of allowed mass splitting among scalar Loryons in \cref{tab:scalarsparam}. This is expressed in terms of the parameter $r_\text{split}$, a rescaling of $\lambda_{h\Phi}^\prime$ defined in \cref{subsec:reps}. \cref{tab:scalarScoreS} shows the scalar Loryons which remain viable after applying the constraint from the $S$ parameter.

For \textbf{fermions}, the situation is more complicated. The Yukawa interaction couples two different representations, so even if there is no vector-like mass and no mass splitting, there can be a significant contribution to $S$. It is straightforward to interpret the constraints in terms of $f_\text{max}$, and this is presented in~\cref{tab:fermionsparam}. There are generally a number of large individual contributions to $S$ which manage to cancel each other. These possibilities can then be ruled out by other constraints. If the lightest mass eigenvalue coupled to the Higgs is less than half the Higgs mass, the scenario is ruled out by Higgs decay constraints. If the lightest mass eigenvalue is greater than half the Higgs mass, the constraint comes from $\k_\g$. This motivates also presenting the results in terms of $\sum\h fQ^2$, to make the interplay with constraints on $\k_\g$ clear.\footnote{Note that this is in fact an underestimate of the contribution to $\k_\g$, since the constraint on $\sum\h fQ^2$ is for asymptotically heavy particles and lighter particles give a larger contribution than asymptotically heavy ones. Properly considering the effect of particles which are not asymptotically heavy on $\k_\g$ does not meaningfully extend the bound past the region excluded by Higgs decays. }  Some examples of the range of allowed $f_\text{max}$ for fermionic Loryons are summarized in \cref{tab:fermionsparam}, and the scorecard showing those fermionic Loryons which remain viable is presented in \cref{tab:fermionScoreS}.

\section{Direct Search Constraints}
\label{sec:direct}
We now turn to consider constraints from direct searches on the Loryon candidates that remain viable when confronted by the indirect constraints studied above. In contrast to indirect bounds, direct bounds are strongly sensitive to Loryon couplings to SM particles other than the Higgs as these interactions typically govern the final state that would be observed at colliders. Our aim here is not to consider all possible direct limits on all possible spectra and couplings but rather to understand the qualitative parameter space allowed by direct searches under generic assumptions.

As noted in \cref{sec:candidates}, we have restricted our attention to Loryon candidates whose hypercharge assignments allow all BSM charged particles to decay into SM final states through either marginal or irrelevant interactions. This is because heavy stable charged particles (HSCPs) are strongly constrained by LHC searches for anomalous ionization energy loss and time of flight, with current bounds above the TeV scale for HSCPs carrying a range of quantum numbers \cite{Khachatryan:2016sfv, Aaboud:2018hdl, Aaboud:2019trc}. Although bounds are somewhat weaker on scalars carrying only hypercharge (around 430 GeV per \cite{Aaboud:2019trc}), these limits are still considerably stronger than the corresponding limits on their promptly-decaying counterparts.

For Loryons neutral under $SU(2)_L$, we identify the lowest-dimensional operators that would allow such decays and assume that (1) all allowed decay operators of the lowest nontrivial dimension are present, and (2) Loryon decays are dominated by the combination of these operators giving the weakest bound. We do not require the decay couplings to respect custodial symmetry as these couplings may be numerically quite small -- consistent with bounds on custodial symmetry violation -- while still allowing for prompt decays. In the same spirit, we assume the decay couplings are small enough that they do not provide significant new production modes.  In many cases, the leading operators carry SM flavor indices; we assume the flavor structure is such that strong flavor-dependent constraints (from e.g.~flavor-changing neutral currents or proton decay) are avoided. In many cases, the bound on a given Loryon candidate depends on the flavor composition of the final state; in quoting a limit we highlight the flavor structure that results in the weakest limit.

For Loryons charged under $SU(2)_L$, there are typically one or more electrically neutral particles in the multiplet that may be the lightest mass eigenstate. In this case, the charged components of the multiplet can decay into the neutral component and SM bosons without assuming any additional couplings. This leads to a missing energy signature whose strength depends sensitively on the mass spectrum of the new particles, with large splittings leading to correspondingly larger (and better-constrained) missing energy signals. In such cases, the bounds on $SU(2)_L$-charged Loryons are typically weakest if no additional interactions are assumed beyond the irreducible couplings to the Higgs.

For scalar Loryons, the bounds from precision electroweak constraints can be avoided by minimizing the mass splitting within a given electroweak multiplet, in which case gauge eigenstates are also approximate mass eigenstates. As such, in determining the state of direct limits on scalar Loryons, it suffices to consider searches for distinct $SU(3)_c \times SU(2)_L \times U(1)_Y$ representations. An approximate direct limit on scalar Loryons in a given custodial representation can then be found by stacking the limits on the $SU(3)_c \times SU(2)_L \times U(1)_Y$ representations that compose the custodial multiplet.

For fermionic Loryons, electroweak symmetry breaking mixes components of different gauge eigenstates, so that gauge eigenstates are no longer approximate mass eigenstates. Moreover, as a custodial multiplet of fermionic Loryons necessarily contains a field charged under $SU(2)_L$, the strength of direct search limits depends sensitively on the mass splitting between the lightest neutral fermion and heavier fermions, which controls the amount of missing energy in the final state. As such, obtaining the direct limit on fermionic Loryons in a given custodial multiplet requires reinterpreting the relevant searches in the space of couplings for that multiplet. We first present the simpler direct limits on scalar Loryons before treating the more complicated direct limits on fermionic Loryons.

\subsection{Scalar Loryons}

The state of direct search limits on the $SU(3)_C \times SU(2)_L \times U(1)_Y$ representations of states that appear as components of the viable custodial multiplets of scalar Loryons is summarized in \cref{tab:directlims}, subject to the above considerations.

\begin{table}[h]
\renewcommand{\arraystretch}{1.6}
\setlength{\tabcolsep}{0.7em}
\setlength{\arrayrulewidth}{1.2pt}
\small
\begin{center}
\begin{tabular}{l|l|c|c|c}
Field & Charge & Decay Couplings & Limit & Ref.  \\ \hline \hline
$S$ & $(1,1)_0$ & -- or $S \times \mathcal{O}^\text{SM}_4$  & -- & \cite{Craig:2014lda} \\ \hline
$S_1$ & $(1,1)_1$ & $S_1^\dag \bar L_i i \sigma_2 L^c_j$  & $\sim 325$ GeV & \cite{Crivellin:2020klg}  \\ \hline
$\omega_1$ & $(3,1)_{-1/3}$ & $\omega_1^\dag \left( \bar Q i \sigma_2 Q^c + \bar d u^c \right)$  & $\sim 520$ GeV & \cite{Sirunyan:2018rlj} \\ \hline
$\omega_2$ & $(3,1)_{+2/3}$ & $\omega_2^\dag \bar d d^c$ & $\sim 520$ GeV & \cite{Sirunyan:2018rlj} \\ \hline
$\Phi_1$ & $(1,2)_{1/2}$ & $-$ (if inert) & $\sim 70$ GeV &\cite{Dercks:2018wch} \\ \hline
 $\Xi_0$ & $(1,3)_0$ & Mixing via $\Xi^a H^\dagger \sigma^a H$ & $\sim 230$ GeV & \cite{Bell:2020gug} \\ \hline
& & $-$ (if inert) & $\sim 275$ GeV & \cite{Chiang:2020rcv} \\ \hline
$\Phi_3$ & $(1,2)_{3/2}$ & $(\Phi_3^\dagger) H \bar d u$ & $\sim 80$ GeV & \cite{Abbiendi:2013hk} \\ \hline
$\Xi_1$ & $(1,3)_1$ & $\Xi_1^{I} [\sigma^I \epsilon]_{\alpha\beta} H^\dagger_\alpha H^\dagger_\beta$ & $\sim 350$ GeV & \cite{Aad:2021lzu} \\ \hline
$\Theta_1$ & $(1,4)_{1/2}$ & $(\Theta_1^\dagger)_{(abc)} H^a H^b \tilde H^c$ & $\gtrsim 350$ GeV & \cite{Aad:2021lzu} \\ \hline
$\Theta_3$ & $(1,4)_{3/2}$ & $(\Theta_3^\dagger)_{(abc)} H^a H^b H^c$ & $\gtrsim 350$ GeV & \cite{Aad:2021lzu}
\end{tabular}
\end{center}
\caption{Assumed decay couplings and direct search limits on scalar Loryons organized by SM representation.\label{tab:directlims}}
\end{table}%

The neutral scalar $S$ need not possess additional couplings that would allow it to decay, leading to a missing energy signature if it remains stable on detector length scales. We assume that it does not couple linearly to invariants constructed purely from the Higgs doublet $H$ as this would lead to mass mixing with correspondingly tighter constraints. This leaves open the possibility that $S$ may decay through dimension-5 operators in which $S$ couples to dimension-4 gauge-invariant SM operators $\mathcal{O}_4^\text{SM}$, excluding $\mathcal{O}_4^\text{SM} = |H|^4$. When $m_S < m_h/2$, it may be produced in the decay of on-shell Higgs bosons and is subject to the constraints on the Higgs invisible or BSM width discussed in \cref{sec:higgs}; the strength of these constraints depends on the fraction $f_\text{max}$ of mass-squared that $S$ acquires from the Higgs. As we account for these bounds in terms of Higgs coupling measurements rather than direct searches, we omit them in \cref{tab:directlims}. For $m_S \geq m_h/2$, $S$ is produced via off-shell Higgs bosons with a modest rate that remains essentially unconstrained by missing energy searches \cite{Craig:2014lda}. The low production rate is such that prompt decays through a variety of SM operators $\mathcal{O}_4^\text{SM}$ are likewise unconstrained. The HL-LHC with 3/ab is not expected to attain sensitivity to off-shell production for $0.5 \leq f_\text{max} \leq 1$.

The hypercharged scalar $S_1$ admits a marginal coupling to two lepton doublets, which is antisymmetric in flavor space due to the antisymmetry of the $SU(2)_L$ contraction. The quoted bound on $S_1$ in \cref{tab:directlims} is obtained in \cite{Crivellin:2020klg} from the reinterpretation of LHC slepton searches, most notably a $\sqrt{s} = 13$ TeV ATLAS analysis with 139/fb of data \cite{Aad:2019vnb}. The detailed mass reach depends on the relative branching ratios into different lepton flavors; it is maximized at $m \sim 325 \text{ GeV}$ when ${\rm BR}(S_1 \rightarrow e^+ \nu) = 0.5 (0), {\rm BR}(S_1 \rightarrow \mu^+ \nu) = 0 (0.5)$ and minimized at $m \sim 200 \text{ GeV}$ when ${\rm BR}(S_1 \rightarrow e^+ \nu) = {\rm BR}(S_1 \rightarrow \mu^+ \nu) = 0.25$ using only the same-flavor bins (note that the sum of branching ratios into electrons and muons can never be less than 50\%). However, we note that the ATLAS search also includes different-flavor bins, although these are not used in the single-slepton interpretation. These bins have comparable sensitivity to the same-flavor bins, and so we expect the limit from opposite-flavor final states to be comparable to those from same-flavor final states. We thus take the limit to be $m \sim 325 \text{ GeV}$ regardless of the relative branching ratios into electrons and muons. The projected HL-LHC bound with 3/ab is expected to reach $m \sim 400 \text{ GeV}$ under the same assumptions \cite{Crivellin:2020klg}.

The colored and hypercharged scalars $\omega_{1}$ and $\omega_{2}$ admit marginal couplings allowing them to decay. The $\omega_{1}$ has the quantum numbers of a leptoquark and can couple to SM fermion bilinears involving one lepton and one quark, as well as SM fermion bilinears consisting solely of quarks. Given the relatively stronger bounds on leptonic decays of colored particles, we assume that the branching ratios into quarks dominate. In contrast, the $\omega_{2}$ only admits a marginal coupling to pairs of down-type quarks. Assuming that decays into quarks dominate, the quoted bound on $\omega_{1}$ and $\omega_{2}$ in \cref{tab:directlims} comes from a $\sqrt{s} = 13$ TeV CMS search with 36/fb for pair-produced resonances decaying to pairs of quarks \cite{Sirunyan:2018rlj}, which excludes stops decaying into light-flavor quarks up to $m\sim 520$ GeV. We are currently unaware of projections for the performance of this search at the HL-LHC.

The electroweak doublet scalar $\Phi_1$ extends the Higgs sector to a two Higgs doublet model, with a range of signatures and constraints depending on the parameters of the potential and couplings to fermions. However, a simple irreducible limit may be obtained by treating $\Phi_1$ as inert, forbidding renormalizable couplings to fermions by imposing the discrete symmetry $\Phi_1 \leftrightarrow - \Phi_1$ and assuming the potential is such that $\Phi_1$ does not acquire a vev. Although $\Phi_1$ admits a range of marginal couplings that would allow it to decay, these may be forbidden by the discrete symmetry. The lightest mass eigenstate is typically a neutral scalar, which can be stable on detector length scales. The charged components of $\Phi_1$ are split from the lightest neutral component by both tree-level and one-loop effects; the former effects are bounded by perturbativity constraints, while the latter lead to splittings on the order of $\sim 350$ MeV. This leads to an experimentally challenging scenario in which the charged-neutral mass splitting is too small to produce distinctive decay products and significant missing energy, but too large to generate a long disappearing track. We take the inferred LEP bound of $m_{H^\pm} > 70$ GeV appearing in \cite{Dercks:2018wch}.

The electroweak triplet, hypercharge-neutral scalar $\Xi_0$ admits a marginal coupling $\Xi_0^a H^\dag \sigma^a H$ that allows it to decay via mixing with the Higgs after electroweak symmetry breaking; this mixing can be sufficiently small to allow for prompt decays without running afoul of indirect constraints. In this case, \cite{Bell:2020gug} obtained a limit of $m \sim 230$ GeV by the reinterpretation of $\sqrt{s} = 13$ TeV ATLAS \cite{Aaboud:2017nhr} and CMS \cite{Sirunyan:2017lae} multi-lepton searches with 36/fb, noting that a naive extrapolation to results of the full Run 2 data set were expected to improve the bound to $m \sim 330$ GeV. Alternately, mixing may be forbidden by imposing the discrete symmetry $\Xi_0 \leftrightarrow -\Xi_0$, rendering it inert and the neutral component stable. In this case, the $\sim 160$ MeV radiative splitting between the charged and neutral components gives rise to a disappearing track signature, leading \cite{Chiang:2020rcv} to obtain a bound of $m \sim 275$ GeV by reinterpreting a $\sqrt{s} = 13$ TeV ATLAS search with 36/fb \cite{Aaboud:2017mpt}. An HL-LHC bound of $\sim 520$ GeV was projected by \cite{Chiang:2020rcv}.

The electroweak doublet, hypercharge-$3/2$ scalar $\Phi_3$ does not admit marginal couplings that would allow its doubly-charged or singly-charged components to decay. Rather, they may decay via operators coupling $\Phi_3$ to a Higgs boson and two fermions \cite{Rentala:2011mr}. If the dominant decay coupling involves leptons, strong constraints from same-sign dilepton searches imply $m \gtrsim 700-900$ GeV depending on the flavor composition of the leptonic branching ratios \cite{Aaboud:2017qph}. If the dominant decay coupling involves quarks, however, the electroweak production cross section is too small to be meaningfully constrained by the CMS paired dijet resonance search \cite{Sirunyan:2018rlj}. Instead, the leading bound arises from LEP searches for hadronically decaying singly-charged Higgses, of order $m \sim 80$ GeV \cite{Abbiendi:2013hk}.

The electroweak triplet, hypercharge-1 scalar $\Xi_1$ admits marginal couplings to either two same-sign leptons or two Higgs bosons. The former leads to a strong bound of $m \gtrsim 700-900$ GeV from same-sign dilepton searches \cite{Aaboud:2017qph}, while the latter leads to a somewhat weaker constraint $\sim 350$ GeV from same-sign $WW$ signals in multi-lepton searches \cite{Aad:2021lzu}. The situation is similar for the electroweak quartet, hypercharge-$1/2$ or $3/2$ scalars $\Theta_1$ and $\Theta_3$. Both admit renormalizable couplings to three Higgs bosons that lead to same-sign $WW$ production similarly constrained by multi-lepton searches \cite{Aad:2021lzu}. Note that the higher-dimensional $SU(2)_L$ representations have correspondingly larger cross sections, which we do not account for here; this has a modest effect on the mass limit.

\subsection{Fermionic Loryons\label{sec:directfermion}}

As noted above, the significant mass mixing among different gauge eigenstates of fermionic Loryons in a given custodial multiplet requires a more detailed treatment of direct search limits. For each of the fermionic Loryon candidates that remains viable after imposing indirect bounds, mass mixing leads to a split mass spectrum and qualitatively similar collider signatures: heavy mass eigenstates are pair produced via Drell-Yan before decaying into light mass eigenstates via the emission of $W, Z$, and Higgs bosons. The light mass eigenstates typically include both charged and neutral fermions that are degenerate at tree-level in the limit of exact custodial symmetry, but one-loop corrections induce small splittings that allow the light charged fermions to decay into their neutral counterparts plus soft SM particles. If the splittings are sufficiently small, the light charged fermions give rise to a distinctive disappearing track signature. If the splittings are somewhat larger, above about 400 MeV, the disappearing tracks are too short to be picked up in existing searches. For fermionic Loryons in larger custodial representations, the spectrum includes additional mass eigenstates intermediate between the light and heavy states. These can lead to two-step cascade decays with higher multiplicities of SM vector bosons in the final state.

To determine limits on fermionic Loryons coming from the decay of heavy mass eigenstates into lighter ones via $W,Z$, and $h$, we reinterpret a series of ATLAS and CMS electroweakino searches at $\sqrt{s} = 13$ TeV, namely the CMS search for two oppositely charged same-flavor leptons and missing transverse momentum \cite{Sirunyan:2020eab} and the ATLAS search for three leptons from on-shell $W,Z$ bosons plus missing transverse momentum \cite{Aad:2019vvi}, both of which are sensitive to pair production events in which one heavy eigenstate decays via a $W$ boson and the other via a $Z$ boson; the ATLAS \cite{Aad:2019vvf} and CMS  \cite{CMS:2021lzg} searches for final states with one lepton, missing transverse momentum, and a Higgs decaying to $b \bar b$, are sensitive to pair production events in which one heavy eigenstate decays via a $W$ boson and the other via a Higgs boson; and the ATLAS search for two leptons and missing transverse momentum \cite{Aad:2019vnb} is sensitive to pair production events in which both heavy eigenstates decay via a $W$ boson. For each of these searches, the collaborations present excluded cross sections for an exclusive channel as a function of the heavy and light fermion masses, allowing for straightforward reinterpretation. Although both collaborations also pursue searches sensitive to pair production events with $hh, ZZ,$ or $hZ$ plus missing transverse momentum, these limits are presented as a function of the heavy mass assuming a (small) fixed value of the light mass, so we are unable to reinterpret these constraints in the full parameter space of interest. The reach of these searches is comparable to the corresponding limits of the searches we reinterpret. For larger custodial representations with additional mass eigenstates intermediate between the light and heavy states, we have estimated the sensitivity of ATLAS and CMS multi-lepton searches \cite{CMS:2019lwf, ATLAS:2021eyc} to the resulting multi-boson final states; we find that they do not significantly improve the limits set by searches for single-step decays into di-boson final states.

In addition, we must consider possible limits on the light mass eigenstates coming from disappearing track searches, which are controlled by the splitting between the lightest charged and neutral fermions. If the splitting is sufficiently small, the light charged fermions become long-lived, and the resulting disappearing track signature may provide an even stronger constraint on the parameter space. For the custodial multiplets under consideration, the splitting between the charged and neutral fermions depends on possible soft breaking of custodial symmetry at tree level, as well as two different types of one-loop corrections. At tree level, it is possible to softly break custodial symmetry by introducing different vector-like mass terms for the Standard Model representations within a given custodial multiplet. For example, in the $[2,1]_{1/2} \oplus [1,2]_{1/2}$ model, differences in the vector-like masses of the $E_0$ and $E_1$ fields comprising the $[1,2]_{1/2}$ custodial irrep will split the masses of the lightest charged and neutral fermions. In principle, a small soft breaking of custodial symmetry can always be introduced to generate a sufficiently large charged-neutral mass splitting to avoid constraints from disappearing track searches without significantly worsening precision electroweak constraints.

However, it is perhaps more compelling to assume exact custodial symmetry at tree level and consider the two types of irreducible splitting due to radiative corrections. The first of these is the familiar finite radiative correction due to the Coulomb energy stored in the charged fields \cite{Thomas:1998wy, Cirelli:2005uq}. The second is the logarithmically divergent radiative correction due to the breaking of custodial symmetry by hypercharge. The latter effect can be straightforwardly computed from \eg~the renormalization group evolution of the Loryons' vector-like masses and Yukawa couplings proportional to the hypercharge gauge coupling. For the $[2,1]_{1/2} \oplus [1,2]_{1/2}$ and $[1,3]_0 \oplus [2,2]_0$ models, the resulting one loop splitting between the lightest charged and neutral mass eigenvalues $m_1^\pm, m_1^0$ at leading logarithmic order is given by
\begin{align}
m_1^\pm - m_1^0 \simeq \frac{3\s g'^2}{16 \pi^2}\s m_1^0 \log\frac{\Lambda}{\mu}\,,
\end{align}
where $\Lambda$ is a UV scale at which custodial symmetry is presumed to be exact and $\mu$ is the renormalization scale. The logarithmic enhancement of this effect causes it to dominate over the finite correction and for $\Lambda \gtrsim$ TeV yields a mass splitting large enough to evade current limits from disappearing track searches \cite{ATLAS:2021ttq}. As such, the direct search limits on fermionic Loryons remain dominated by the decays of heavy mass eigenstates into light ones, even before including possible tree-level soft breaking of custodial symmetry.

\begin{figure}[t!]
\includegraphics[width=\textwidth]{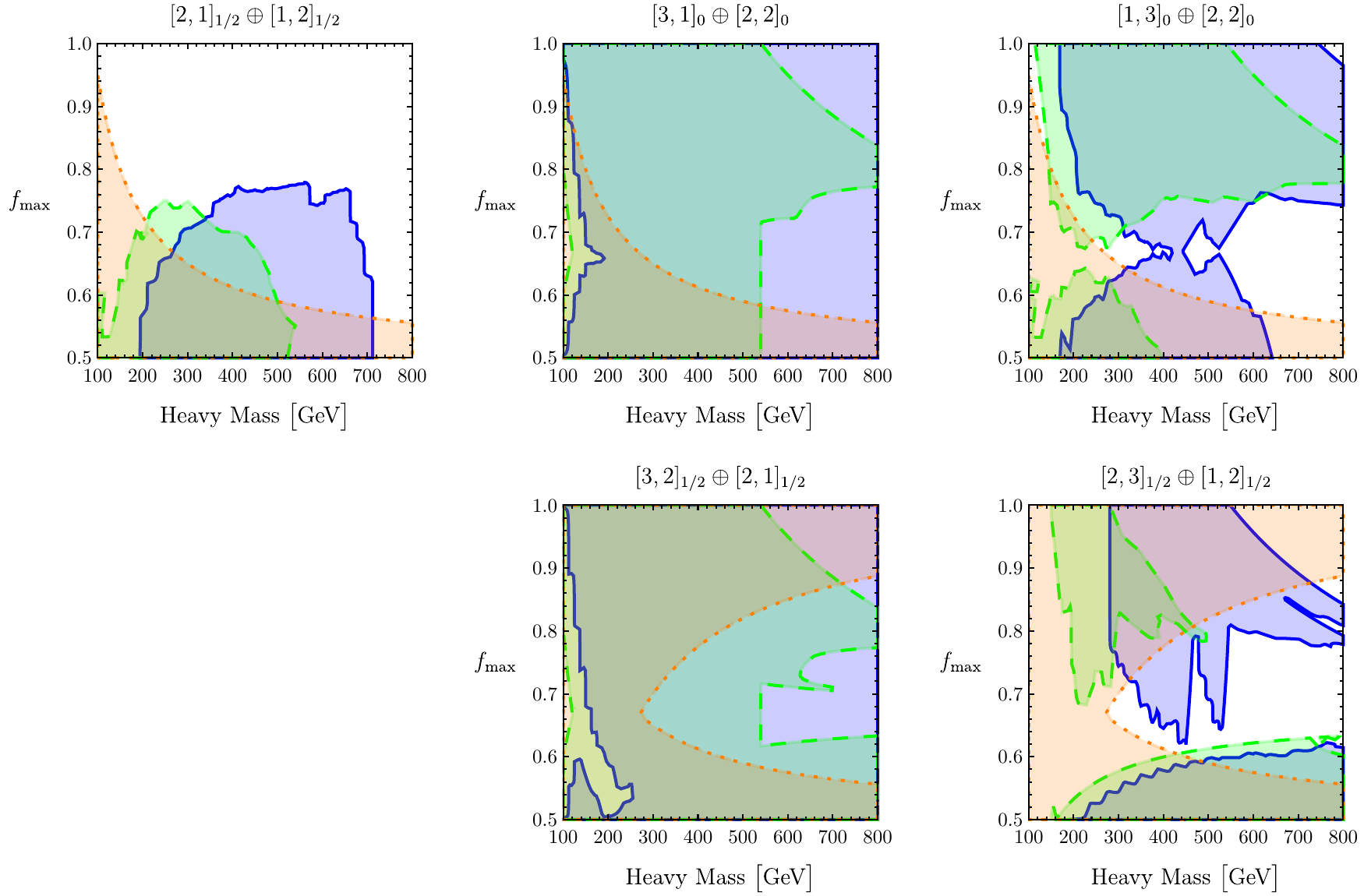}
\caption{Exclusions on fermionic Loryons from LHC searches for missing energy plus $WZ$ (blue, solid) or $WW$ (green, dashed) bosons, along with the the LEP II bound (orange, dotted) on charged fermions, as a function of the mass of the heaviest fermions and the fraction $f_\text{max}$ of their mass which is Higgs-dependent. Only combinations of custodial representations that remained viable after imposing the constraints in previous sections are shown. \label{fig:directlims}}
\end{figure}

In order to determine the bound on fermionic Loryons in a given custodial multiplet from these searches, we compute the leading-order production cross sections and branching ratios for the heavy mass eigenstates in a given multiplet using \texttt{FeynRules} \cite{Alloul:2013bka}, \texttt{FeynArts} \cite{Hahn:2000kx}, and \texttt{FormCalc} \cite{Hahn:1998yk, Hahn:2016ebn} and compare the relevant production cross sections times branching ratios to the excluded cross section in each search. In doing so, we include decays into both charged and neutral fermions among the light mass eigenstates as the charged mass eigenstates subsequently decay into the neutral mass eigenstates plus additional soft particles that should not significantly impact the acceptance of the searches in question. We further include the combined LEP bound on charged fermions  \cite{lepsusy}, taking the limit to be $m^\pm \gtrsim 90$ GeV in light of the possible variations in the charged-neutral mass splitting \cite{Egana-Ugrinovic:2018roi}. The results are summarized in \cref{fig:directlims}.

The parametric behavior of the limits in \cref{fig:directlims} can be most clearly understood for the $[2,1]_{1/2} \oplus [1,2]_{1/2}$ model shown in the first panel. The LHC searches are most sensitive when the splitting between heavy and light mass eigenstates is large, for this leads to the largest amount of missing energy. In the parameter space of \cref{fig:directlims}, this corresponds to smaller values of $f_{\rm max}$, for which mass splittings are induced by competition between EWSB-dependent and -independent contributions. At lower values of the heavy mass, the available phase space for producing electroweak bosons begins to close off, while at higher values of the heavy mass, the production cross section falls off. The shape of the LEP exclusion in this parameter space is set by the mass of the light mass eigenstates, which include a charged Dirac fermion; at lower values of $f_{\rm max}$ the large mass splitting drives the light mass eigenstates below the LEP bound, and less splitting is required to reach the LEP bound as the heavy mass decreases. The shape of the limits for larger custodial representations in the remaining panels of \cref{fig:directlims} is governed by the same logic but no longer takes a simple form in the plane of $f_{\rm max}$ and the heavy mass. This is because the higher-dimensional custodial representations lead to more than two clusters of mass eigenstates, so direct search limits are most sensitive to the splitting between intermediate and light mass eigenstates. The precise values of these splittings and the electromagnetic charges of the light eigenstates varies from multiplet to multiplet, leading to the observed pattern of exclusion regions.

\begin{table}[t!]
\renewcommand{\arraystretch}{1.6}
\setlength{\tabcolsep}{0.7em}
\setlength{\arrayrulewidth}{1.2pt}
\centering
\begin{tabular}{  c | c | c | c | c | c | c | c | c  }
\multicolumn{9}{c}{\textsc{Vector-like Fermion Scorecard}}\\[5pt]
 & $R=1$ & 2 & 3 & 4 & 5 & 6 & 7 & 8 \\
\hline
$L=1$ & $|Y_\text{max}|=\times$ & $\sim\!\frac12$ & $\sim\!0$ & $\times$ & $\times$ & $\times$ & $\times$ & $\times$ \\
\hline
2 & $\sim\!\frac12$ & $\sim\!0$ & $\times$ & $\times$ & $\times$ & $\times$ & $\times$ & $\times$ \\
\hline
3 & $\times$ & $\times$ & $\times$ & $\times$ & $\times$ & $\times$ & $\times$ & $\times$ \\
\hline
4 & $\times$ & $\times$ & $\times$ & $\times$ & $\times$ & $\times$ & $\times$ & $\times$ \\
\hline
5 & $\times$ & $\times$ & $\times$ & $\times$ & $\times$ & $\times$ & $\times$ & $\times$ \\
\hline
6 & $\times$ & $\times$ & $\times$ & $\times$ & $\times$ & $\times$ & $\times$ & $\times$ \\
\hline
7 & $\times$ & $\times$ & $\times$ & $\times$ & $\times$ & $\times$ & $\times$ & $\times$ \\
\hline
8 & $\times$ & $\times$ & $\times$ & $\times$ & $\times$ & $\times$ & $\times$ & $\times$
\end{tabular}\\[13pt]
\caption{The representations for the vector-like fermionic BSM Loryons still viable after considering direct search constraints. A $\sim$ means the representation is not viable on its own but can be added together with other BSM representations to flip the sign of $\kappa_\gamma$.}
\label{tab:fermionScoreDirect}
\end{table}

\section{Viable Loryons}
\label{sec:viable}
\begin{figure}[t!]
\includegraphics[width=\textwidth]{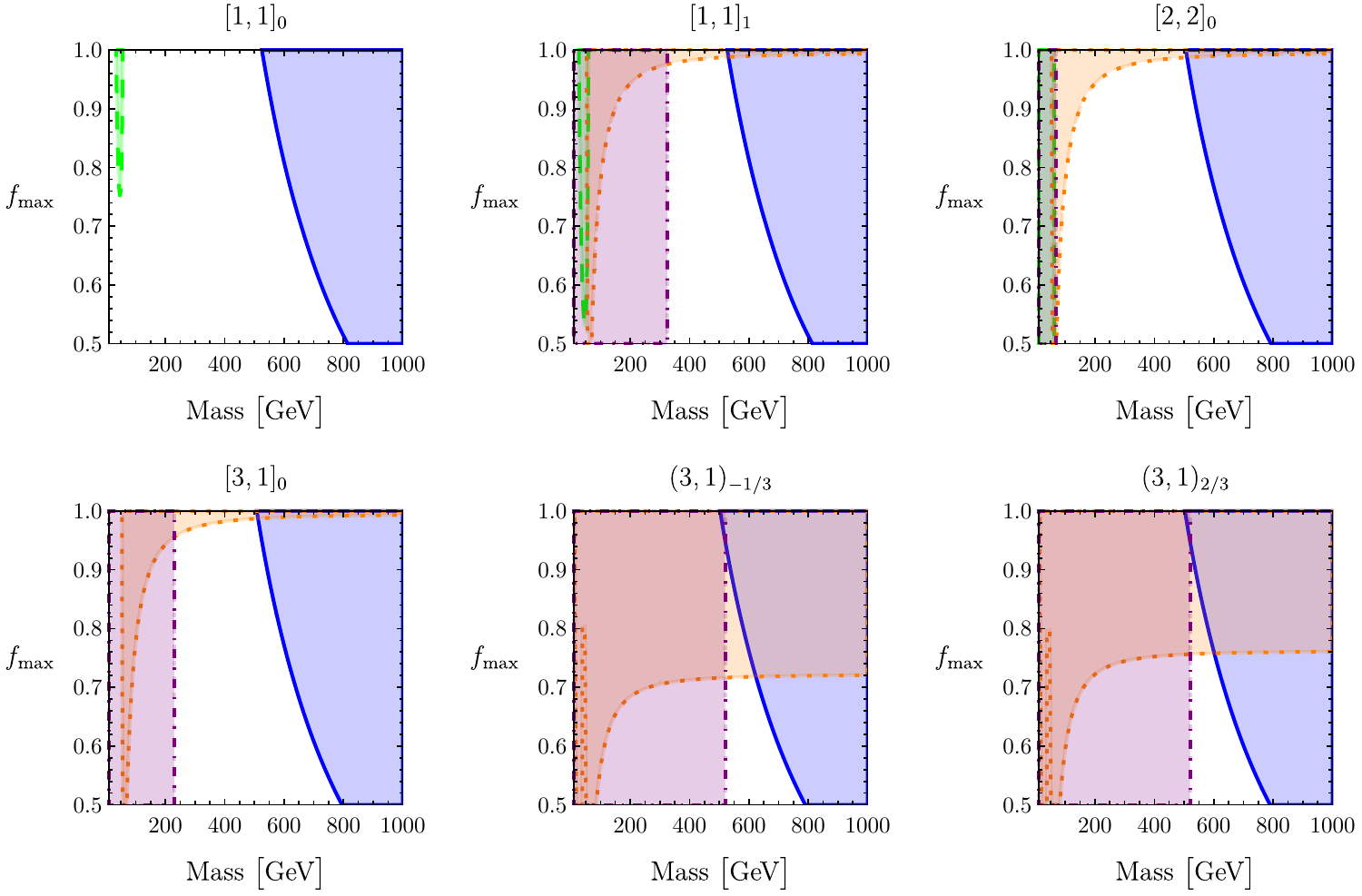}
\caption{Regions of parameter space for scalar Loryons which are ruled out by one of the constraints described earlier. The orange, dotted region is ruled out by constraints on $\k_\g$ or $\k_g$; the blue, solid region is ruled out by unitarity bounds; the green, dashed region is ruled out by constraints on Higgs decay; and the purple, dot-dashed region is ruled out by direct search bounds. All plots assume no mass splitting between the components of the Loryon multiplet, which is a non-trivial assumption for $[2,2]_0$. Note that the bottom middle and right plots are indicated by their SM gauge quantum numbers. \label{fig:scalarcombo}  }
\end{figure}
\begin{figure}[t!]
\includegraphics[width=\textwidth]{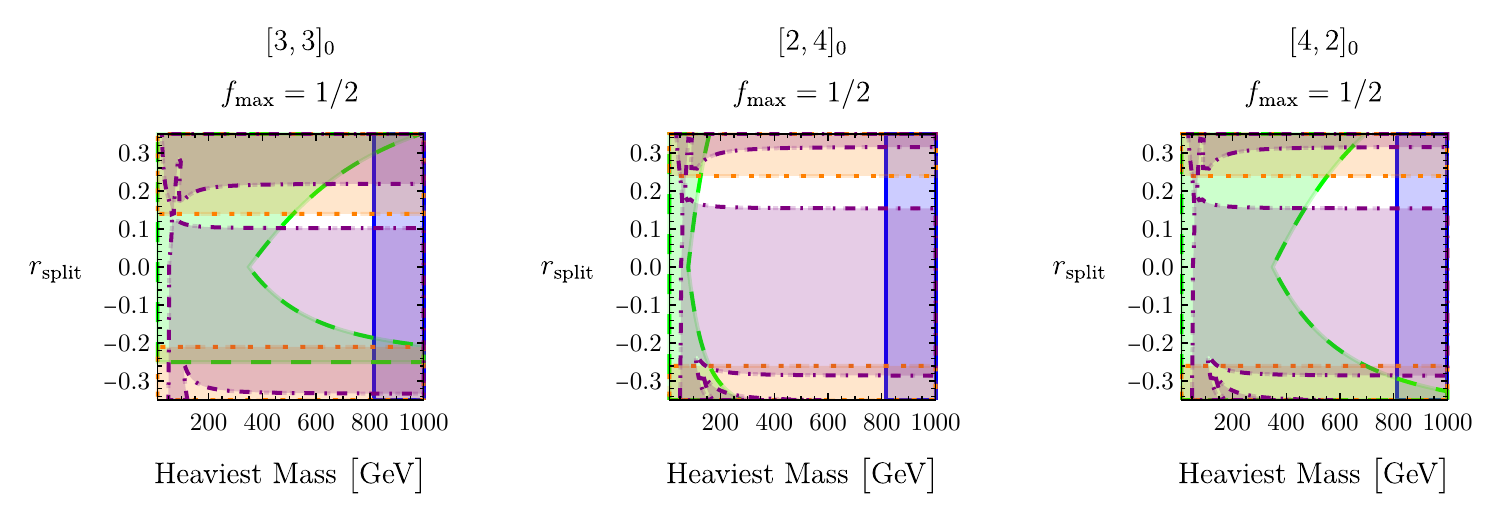}
\includegraphics[width=\textwidth]{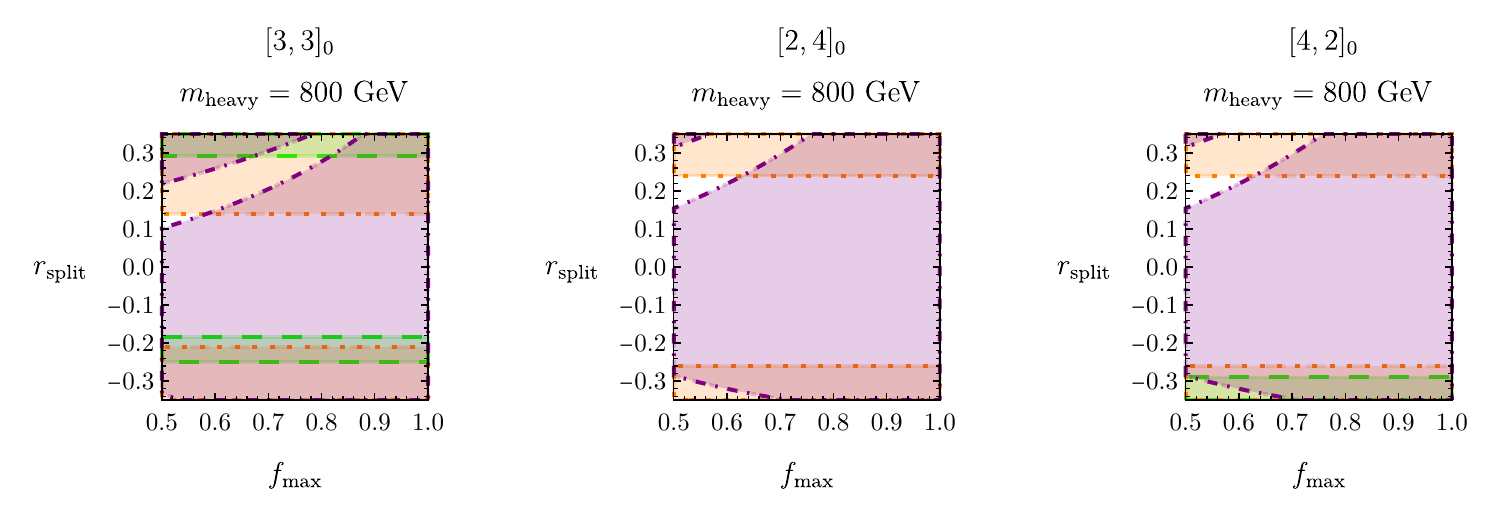}
\caption{Regions of parameter space for scalar Loryons which are ruled out by one of the constraints described earlier. The orange, dotted region is ruled out by the electroweak precision parameter $S$; the purple, dot-dashed region is ruled out by constraints on $\k_\g$; the blue, solid region is ruled out by unitarity bounds; the green, dashed region is ruled out by direct search bounds. The first row has fixed $f_\text{max}=.5$; the second row has fixed the heaviest mass eigenvalue at $800$ GeV.
\label{fig:scalarcombosplitting}}
\end{figure}
\begin{figure}[t!]
\centering
\includegraphics[width=0.7\textwidth]{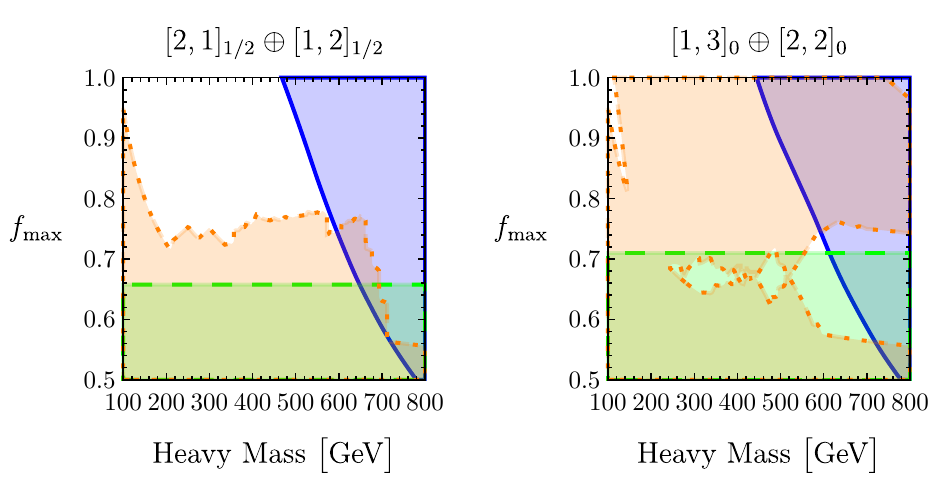}
\caption{Regions of parameter space for fermionic Loryons which are ruled out by one of the constraints described earlier. The blue, solid region is ruled out by unitarity bounds; the green, dashed region is ruled out by precision electroweak measurements; the orange, dotted region is ruled out by direct search constraints. Note that we do not include constraints from Higgs couplings, which nominally rule out all fermionic Loryon candidates in combination with precision electroweak measurements; constraints from $\k_\g$ may only be satisfied by including some additional number of scalar Loryons or non-Loryons coupling to the Higgs. \label{fig:fermioncombo}}
\end{figure}

Putting everything together, we may now address the animating question of this work: does current data allow for Loryons beyond the Standard Model? Surprisingly, a number of Loryon candidates remain viable in light of direct and indirect limits, although the situation is far more optimistic for scalar Loryons relative to their fermionic counterparts.

The current status of \textbf{scalar} Loryons is summarized in \cref{fig:scalarcombo} and \cref{fig:scalarcombosplitting}. The parameter space has particularly large regions of viability for the custodial representations $[1,1]_0, [2,2]_0, [1,3]_0,$ and $[3,1]_0$ since the current direct search bounds on the SM singlet $S_0$, the charged singlet $S_1$, the bi-doublet $\Phi_1$, and the hypercharge-neutral triplet $\Xi_0$ vary from essentially nonexistent to $\sim 325$ GeV, while the indirect bounds from Higgs coupling measurements are modest.
Among the scalar Loryons, constraints from $\k_\g$ permit a charge-1 scalar Loryon without flipping the sign of $\k_\g$, but they narrowly exclude charged Loryons acquiring {\it all} of their mass from the Higgs. The custodial symmetry representations $[1,1]_0$, $[2,2]_0$, $[3,1]_0$, and $[1,3]_0$ each include a single charge-1 particle. For the $[1,1]_Y$ representation, constraints on $\k_g$ leave open the possibility for the Loryon to be a color triplet. In addition, scalars in larger custodial representations can remain viable if sufficient mass splitting is introduced so that only a single charge-1 state has a significant coupling to the Higgs; this can be done for the representations $[3,3]_0$, $[2,4]_0$, and $[4,2]_0$ (see \cref{fig:scalarcombosplitting}).  Another avenue by which larger custodial representations can survive the bounds is to give them larger hypercharges; this yields enough charged particles to flip the sign of $\k_\g$. Direct searches place lower bounds on the masses ranging from $\sim$ one hundred to several hundred GeV, while perturbative unitarity places upper bounds on their mass to be at most $\sim 800$ GeV.

The current status of \textbf{fermionic} Loryons is summarized in \cref{fig:fermioncombo}. In contrast with the scalar case, the viable parameter space is much more tightly constrained. As noted above, fermionic Loryons are only viable if the model results in flipping the sign of $\k_\g$. While any individual custodial multiplet is insufficient to achieve this, flipping the sign of $\k_\g$ can occur if there are multiple copies of a given multiplet. However, these additional copies increase the contribution to the $S$-parameter such that the constraints of $\k_\g$ and $S$ cannot be simultaneously satisfied by multiple copies of a given multiplet. We conclude that {\it fermionic Loryons, in isolation, are excluded by current data}. In principle, it is possible to satisfy the constraints from $\k_\g$ and $S$ by adding a set of additional states to flip the sign of $\k_\g$ without running afoul of precision electroweak measurements.

Assuming the $\k_\g$ constraints are satisfied in this way (and assuming these other states do not impact the Loryon phenomenology), there are two combinations of custodial representations which still have viable regions of parameter space:  $[2,1]_{1/2}\oplus[1,2]_{1/2}$ and $[1,3]_0\oplus[2,2]_0$. For fermionic Loryons transforming in the $[2,1]_{1/2} \oplus [1,2]_{1/2}$ representation, direct search bounds and precision electroweak constraints both exclude regions where the Higgs-independent contributions to fermion masses are comparable in size to the Higgs-dependent contributions. In this limit, there is large splitting among the mass eigenstates, which both increases the contribution to the $S$-parameter and increases the sensitivity of direct searches that require large missing energy. LEP limits place a lower bound on the overall mass scale, while perturbative unitarity places an upper bound. In the remaining viable region for the $[2,1]_{1/2} \oplus [1,2]_{1/2}$ model, the heavy mass eigenstate is between $\sim 100-600$ GeV, and the Higgs-independent contributions to fermion masses are no more than about one-third the size of the Higgs-dependent contributions.

Constraints are tighter still for the $[1,3]_0\oplus[2,2]_0$ model; it is likely that the viable parameter space could be entirely closed by a proper statistical combination of all limits, an exercise beyond the scope of this work. As with the $[2,1]_{1/2} \oplus [1,2]_{1/2}$ model, precision electroweak measurements exclude regions where the Higgs-independent contributions to fermion masses are comparable in size to the Higgs-dependent contributions. However, direct searches exclude disjoint regions where the Higgs-independent contributions to fermion masses are either comparable in size to, or much smaller than, the Higgs-dependent contributions, on account of how these contributions translate into the spectrum of mass eigenstates. In conjunction with LEP and perturbative unitarity bounds, this essentially closes the parameter space for the $[1,3]_0\oplus[2,2]_0$, with the exception of a very small window, see \cref{fig:fermioncombo}.

\section{Future Prospects}
\label{sec:future}

Given the viable parameter space for scalar and fermionic Loryons in light of current data, it is natural to wonder what the prospects might be for discovering or excluding Loryons at the HL-LHC. In particular, we anticipate that the HL-LHC will significantly improve the precision with which Higgs couplings are determined. This improved sensitivity will either lead to increased coverage of the Loryon parameter space or point the way towards a discovery.

For the most part, we use the HL-LHC Higgs coupling projections in \cite{Cepeda:2019klc}. We begin with the channels that already provide nontrivial constraints with current data. Using the projected improvement for $\k_\g$ and $\k_g$ in \cite{Cepeda:2019klc}, we find that color triplet scalar Loryons could be entirely ruled out. None of the color singlet cases would be ruled out by the improvement in $\k_\g$ alone since the constraint is not expected to tighten enough to eliminate a single charge 1 scalar; the increased precision narrows the open parameter space that survives by flipping the sign of $\k_\g$.

The HL-LHC is also projected to make a fairly precise determination of $\k_{Z\g}$. While the overall magnitude of the resulting constraint is expected to be much weaker than the ones derived using $\k_\g$, the interplay of $\k_{Z\g}$ and $\k_\g$ is quite powerful. In particular, the expected constraint on $\k_{Z\g}$ would probe scenarios that currently remain viable by flipping the sign of $\k_\g$ since the relevant parameter space generally also yields a significant contribution to $\k_{Z\g}$. The contribution of viable Loryons to $\k_{Z \g}$ relative to their contribution to $\k_\g$ is shown in \cref{{tab:zgamma}}. Weak singlet scalars and the viable fermions contribute to $\k_{Z\g}$ with opposite sign relative to the SM, while the other viable new scalar Loryons contribute to $\k_{Z\g}$ with the same sign as the SM. The contribution to $\k_{Z\g}$ is too small to flip the sign of $\k_{Z\g}$ and $\k_\g$ simultaneously.  In more complicated models, a combination of some new singlets and fermions and/or some new non-singlet scalars could conspire to have their contributions to $\k_{Z\g}$ cancel while flipping the sign of $\k_\g$.

\begin{table}[t]
\renewcommand{\arraystretch}{1.6}
\setlength{\tabcolsep}{0.3em}
\setlength{\arrayrulewidth}{1.2pt}
\centering
\begin{adjustbox}{tabular=c | c | c | c | c | c | c | c | c | c | c  ,center}
Rep & $[1,1]_1$ & $[3,1]_0$ & $[1,3]_0$ & $[2,2]_0$ & $[3,3]_0$ & $[4,2]_0$ & $[2,4]_0$ & $[2,3]_{-1/2}$ & $[2,1]_{1/2} \oplus [1,2]_{1/2}$ & $[1,3]_0 \oplus [2,2]_0$ \\
\hline
$\frac{\sum\h_iC_i^{Z\g}}{\sum\h_iQ_i^2}$ & $-.12$ & $.38$ & $-.12$ & $.13$ & $.13$ & $.30$ & $-.032$ & $-.008$ & $-.019$ & $-.019$
\end{adjustbox}
\caption{The contribution of viable Loryons to $\k_{Z\g}$ as a ratio to the contribution to $\k_\g$. To maintain $\k_{Z\g}$ within 20 percent of the SM value while flipping the sign of $\k_\g$ requires $|\sum\h_iC_i^{Z\g}/\sum\h_iQ_i^2|<.074$. \label{tab:zgamma}}
\end{table}

Higgs cross section measurements at the LHC are sensitive to the wavefunction renormalization of the physical Higgs scalar $h$, which is expected to become relevant in the HL-LHC era. For Loryons carrying SM quantum numbers, this constraint is typically weaker than the bounds from $\kappa_\g$ and $\kappa_{Z \g}$. However, for the SM singlet $S_0$, this will eventually provide a non-trivial constraint on the parameter space that is complementary to the existing bounds from exotic Higgs decays or partial wave unitarity. To determine the projected HL-LHC sensitivity to this effect, we use the single-operator bound on
\begin{align}
\mathcal{O}_H \equiv \frac{C_H}{\Lambda^2} \frac{1}{2} \left( \partial^\mu |H|^2 \right)^2\,.
\end{align}
The HEPFit collaboration projects a 95\% CL bound of $\Lambda / \sqrt{|C_H|} = 1.4 \text{ TeV}$~\cite{Cepeda:2019klc}. Two comments are in order. First, although the bound is quoted on the Wilson coefficient of a SMEFT operator (which is never the appropriate EFT description for the low-energy effects of Loryons), the fit simply reflects a bound on a common shift to single-Higgs production processes. As such, it may be equivalently interpreted as a limit on the parameter $\kappa_h$ defined via
\begin{align}
\mathcal{L} \supset \kappa_h \times \frac{1}{2} (\partial h)^2 \,,
\end{align}
which uniformly shifts single-Higgs production rates when $h$ is canonically normalized. With this interpretation, the projected 95\% CL HL-LHC limit on $\Lambda / \sqrt{|C_H|}$ translates to
\begin{align}
\kappa_h \in [0.97,1.03]\,.
\end{align}
Second, although integrating out any heavy state typically generates multiple EFT operators and lead to bounds weaker than those expected from single-operator projections, in the case of the SM singlet $S_0$ the only low-energy effects at the LHC are the common shift in production rates noted above and a shift in the di-Higgs rate from radiative corrections to the Higgs potential. As the anticipated bounds on $\kappa_h$ are much stronger than those expected from di-Higgs measurement, the projected single-operator bound on $\mathcal{O}_H$ provides a reasonable approximation of the expected sensitivity of a global fit to the specific low-energy effects of $S_0$.

Unfortunately, the expected HL-LHC precision lies parallel to, and is somewhat weaker than, the bound from perturbative unitarity, so the HL-LHC is unlikely to significantly improve constraints on an individual singlet scalar Loryon in this channel within the regime of perturbative validity. That being said, if there are some number $N$ of singlet scalar Loryons, the HL-LHC bound on $\k_h$ scales with $N$, while the leading contribution to the perturbative unitarity bound is unaffected. Subleading contributions to the perturbative unitarity bound scale with $N$; further study is required to understand if HL-LHC sensitivity may play an interesting role.

\begin{figure}[t!]
\includegraphics[width=\textwidth]{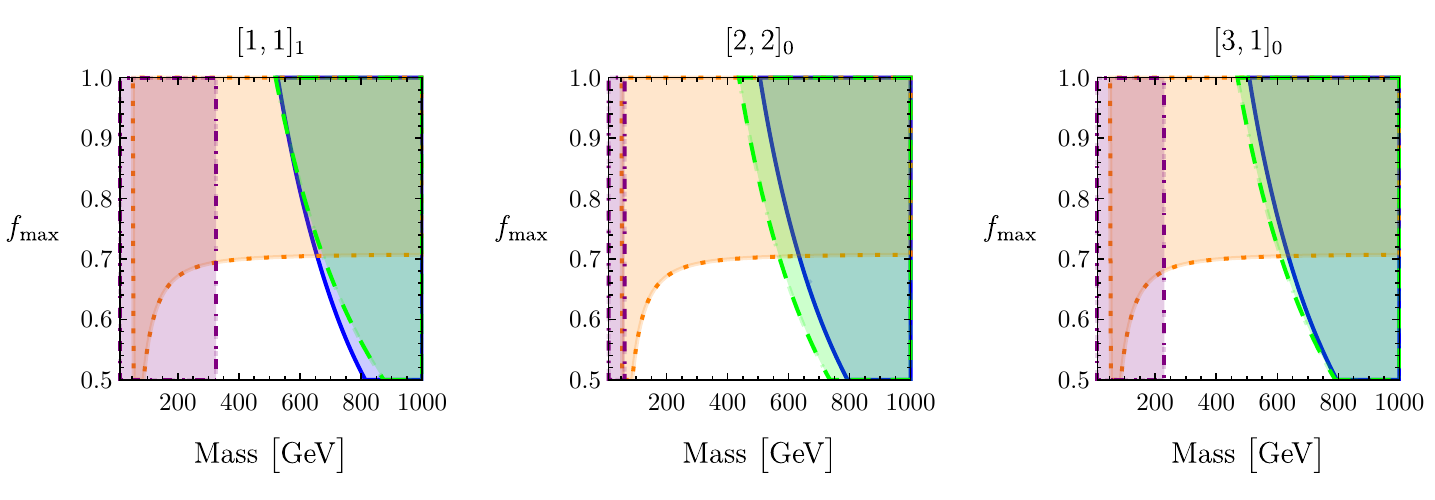}
\caption{Expected sensitivity of the HL-LHC to Loryons based on the Higgs coupling projections in \cite{Cepeda:2019klc}. The orange, dotted region is projected to be ruled out by improved constraints on $\k_\g$ or $\k_g$; the green, dashed region can be ruled out by the HL-LHC measurement of the Higgs cubic; the blue, solid region is ruled out by unitarity bounds; and the purple, dot-dashed region is ruled out by current direct search bounds.}
\label{fig:HLLHCbounds}
\end{figure}

A more promising observable at the HL-LHC is the Higgs self-coupling. Although the anticipated sensitivity of the HL-LHC to deviations in the Higgs self-coupling is significantly less than its sensitivity to wavefunction renormalization effects, the Loryon contribution to the Higgs cubic coupling is enhanced relative to wavefunction renormalization when the Higgs-Loryon coupling is large. Near the perturbative unitarity bound, the correction becomes a $\mathcal{O}(1)$ effect. The leading effect of Loryons on di-Higgs production in the large-coupling limit comes from the direct shift in the coefficient of $h^3$, while subleading effects at large coupling include the above-mentioned shift due to wavefunction renormalization as well as new momentum-dependent contact interactions between $h$ and longitudinal vectors. For simplicity, we consider only the leading effect. As shown in \cref{fig:HLLHCbounds}, the 95\% CL constraint $0.1 < \kappa_\lambda < 2.3$ \cite{Cepeda:2019klc} on modifications to the Higgs self-interaction -- as projected for the HL-LHC -- would provide sensitivity to regions of parameter space otherwise allowed by the perturbative unitarity bound. As the unitarity bound is fixed, more precise measurements of the Higgs self-interaction (such as that expected from the ILC) will have a significant effect on the viable parameter space for new Loryons.

There is also scope for improvement in direct searches for Loryon resonances. Of the scalars with large amounts of viable parameter space remaining in \cref{fig:scalarcombo}, $[1,1]_0$ and $[2,2]_0$ are assumed inert and are difficult to see directly. However, the neutral triplet $[3,1]_0$, if inert, has a projected HL-LHC bound from displaced vertices that rules out most of the viable parameter space -- see \cref{sec:direct} and \cite{Chiang:2020rcv}. If, on the other hand, the neutral triplet has a small mixing with the Higgs, there are many potential dedicated searches for the pair production of (charged or neutral) heavy Higgses, decaying to \eg\ $WW,WZ,tb$, that could fill in the gap \cite{Bell:2020gug}. For the $[1,1]_1$, a dedicated interpretation of the opposite-flavor bins in the existing ATLAS dilepton analysis \cite{Aad:2019vnb} and future analyses would be valuable. For the fermion models, there would be immediate gains from an analysis of neutral diboson signals ($hh,hZ,ZZ$) in the plane of heavy and light fermion masses, analogous to those provided for charged diboson final states that were reinterpreted in \cref{sec:directfermion}.

\section{Conclusions}
\label{sec:conclusion}

Our goal in this work was to classify non-decoupling new particles (``Loryons’’) that are still experimentally viable.  By non-decoupling, we mean that the Loryons acquire at least half of their mass from the Higgs vacuum expectation value; this implies that the low energy effective theory obtained by integrating out the Loryons must be described using HEFT. Taking a mild set of phenomenologically reasonable assumptions, including imposing approximate custodial symmetry and prompt decays of electromagnetically charged particles, the list of possible new states is finite.  We determined the allowed parameter space by imposing partial wave unitarity, along with a set of indirect and direct experimental constraints.  The available parameter space for scalar Loryons remains large, while the fermionic cases are essentially ruled out.

Loryons provide a set of concrete targets for future searches. One can either search for the Loryons directly, or indirectly by interpreting constraints on the HEFT parameter space. In both cases, the HL-LHC stands to significantly improve on existing bounds. Such improvement is strongly motivated as the persistence of Loryons demonstrates that HEFT remains a viable framework for the interpretation of current Higgs data. The exclusion of Loryons would represent significant progress towards verifying that electroweak symmetry can be linearly realized by the known particles. On the other hand, evidence for a Loryon -- above and beyond the thrill of discovery -- would imply that electroweak symmetry is not linearly realized by Standard Model particles on their own.

There are many future directions to explore.  In principle, now that the viable Loryon parameter space is known, it is possible to design new LHC searches to target these interesting regions either directly or indirectly. We discussed many potential new searches/reinterpretations in \cref{sec:future}, but there is considerable room for further development. The complete Loryon parameter space is unlikely to be fully probed at the LHC, and so a dedicated study determining the reach of future colliders is warranted. It would also be interesting to relax some of the assumptions shaping our definition of the Loryon parameter space, which would lead to new signatures. For example, one of the phenomenological requirements made here was that the Loryon decays are all prompt; relaxing this assumption leads to signals in a variety of long-lived particle searches that would open up new pathways to discovery.  We have also assumed that Loryon contributions to flavor- or CP-violating observables are minimized; exploring a wider flavor structure and/or allowing for CP violating couplings would yield many interesting complementary probes of these non-decoupling new particles.

There are also implications for cosmology.  Many of the multiplets include an electrically neutral lightest state, which is an obvious candidate for dark matter.  Some of the Loryons have been studied as dark matter, \eg\ the singlet extension of the Standard Model \cite{Silveira:1985rk, McDonald:1993ex, Burgess:2000yq}, singlet-doublet dark matter \cite{Cohen:2011ec}, or the minimal dark matter program \cite{Cirelli:2005uq}, but others have not been as widely explored.  In general, the relic density of Loryon dark matter candidates is expected to be a rich subject since there are a number of non-trivial allowed couplings between Loryons and the electroweak/Higgs bosons.  Imposing a relic density requirement could be used to further motivate regions of viable Loryon parameter space.  The Loryons are also expected to have an impact on the stability of the Higgs potential, and in fact, one might need to add some additional new physics to (meta-)stabilize the Higgs potential to make these models viable.  Finally, it would be interesting to study the impact of the Loryons on the electroweak phase transition.  We leave the exploration of these connections to cosmology and their impact on the viable Loryon parameter space for future work.

It is remarkable that new particles obtaining most of their mass from electroweak symmetry breaking could still be lurking under our noses.  This paper serves to highlight the fact.  Our hope is that this provides a new set of concrete targets to both motivate new searches at current experiments and augment the case for future colliders.

\acknowledgments

We thank Aleksandr Azatov, Spencer Chang, and Isabel Garcia Garcia for useful conversations, and Zoltan Ligeti for suggesting alternatives to ``Loryon.'' The work of I.~Banta and N.~Craig is supported by the U.S.~Department of Energy under the grant DE-SC0011702. The work of T.~Cohen and X.~Lu is supported by the U.S.~Department of Energy under grant number DE-SC0011640. D.~Sutherland has received funding from the European Union's Horizon 2020 research and innovation programme under the Marie Skłodowska-Curie grant agreement No.~754496. N.~Craig thanks LBNL and the BCTP for hospitality during the completion of this work.

\appendix
\section*{Appendix}
\addcontentsline{toc}{section}{\protect\numberline{}Appendix}%
\section{Effective Lagrangian from Loryons\label{app:quadlags}}

In this Appendix, we calculate the mass spectra of scalar and fermionic Loryons in arbitrary custodial representations, as well as their one-loop corrections to the Higgs effective Lagrangian.

\subsection{Scalars}

In unitary gauge, \cref{eq:HiggsMatrixUnitaryGauge}, $\tr[T^a_2 H T^{\dot a}_2 H^\dagger]= \frac14 (v+h)^2 \delta^{a\dot a}$. The part of the Lagrangian that is quadratic order in $\Phi$ can be written\footnote{This is built from pieces \cref{eq:ScalarMassExplicit,eq:ScalarMassUniv,eq:ScalarMassSplit}, plus a canonically normalized kinetic term.}
\begin{equation}
\mathcal{L}_\text{quad} = - \frac{1}{2^\rho} \tr \left[ \Phi^\dagger \left( D^2 + \frac12 \lambda_\text{ex} v^2 + \frac12 \lambda_{h\Phi} (v+h)^2 \right) \Phi + \lambda^\prime_{h\Phi} (v+h)^2\, \Phi^\dagger T^a_L \Phi T^a_R \right] \,,
\label{eq:QuadraticScalarLagrangian}
\end{equation}
where $D_\mu$ is the covariant derivative.

To elucidate the mass spectrum, we decompose the matrix representation $\Phi$ of the custodial group $SU(2)_L \times SU(2)_R$ into irreps $\phi_V$ of its diagonal subgroup $SU(2)_V$. Here $\phi_V$ are $V$-dimensional vectors built out of linear combinations of the matrix elements $\Phi_{\alpha \dot \alpha}$:
\begin{equation}
\tr \left( U^i \Phi \right) = \bigg[ \underset{V}{\oplus} \phi_V \bigg]^i \,,
\label{eqn:UPhi}
\end{equation}
with
\begin{equation}
V \in \mathcal{V} = \Big\{ L+R-1\;,\; L+R-3\;,\; \cdots \;,\; |L-R|+1 \Big\} \,.
\end{equation}
The explicit coefficients $U^i_{\dot \alpha \alpha}$ are given by the appropriate Clebsch-Gordan coefficients. $U$ satisfies the resolution of the identity
\begin{equation}
\left(U^i_{\dot\beta \beta }\right)^* U^i_{\dot\alpha \alpha} = \delta_{\alpha\beta} \delta_{\dot\alpha\dot\beta} \,,
\label{eq:resident}
\end{equation}
and $U$ is also covariant under arbitrary transformations under the diagonal subgroup:
\begin{equation}
\tr \Big[U \Phi' \Big] = \underset{V}{\oplus}\, \phi'_V
\quad\Rightarrow\quad
\tr \Big[ U \exp(i\epsilon^a T^a_L)\, \Phi\, \exp(-i\epsilon^b T^b_R) \Big] = \underset{V}{\oplus}\, \exp(i\epsilon^a T^a_V)\, \phi_V \,,
\label{eq:DiagonalSubgroupCovariance}
\end{equation}
for $\epsilon^a$ arbitrary. Finding the second order variation of \cref{eq:DiagonalSubgroupCovariance} with respect to $\epsilon^a$ yields
\begin{align}
\tr \left[ U T^{(a}_L T^{b)}_L \Phi + U \Phi T^{(a}_R T^{b)}_R - 2 U T^{(a}_L \Phi T^{b)}_R \right] &= \underset{V}{\oplus}\, T^{(a}_{V} T^{b)}_{V}  \phi_V \,;
\end{align}
contracting with $\delta^{ab}$, and using the fact that $T^a_L T^a_L = C_2(L) \mathbb{1}_L$ in any irrep $L$ then gives
\begin{equation}
  \tr \left[ U T^a_L \Phi T^a_R \right] = \underset{V}{\oplus}\, \frac12 \Big[ C_2(L) + C_2(R) - C_2(V) \Big] \phi_V \,.
\end{equation}
Now inserting a resolution of the identity in \cref{eq:resident}, we get
\begin{align}
\tr \left( \Phi^\dagger T_L^a \Phi T_R^a \right) &= \left[ \tr\left(U \Phi\right) \right]^*\, \tr \left( U T_L^a \Phi T_R^a \right) \,,\notag\\[5pt]
&= \sum_{V}\, \phi_V^\dagger \frac12 \Big[ C_2(L) + C_2(R) - C_2(V) \Big] \phi_V \,.
\end{align}
Therefore, the Lagrangian in \cref{eq:QuadraticScalarLagrangian} decomposes into the sum
\begin{align}
\mathcal{L}_\text{quad} = - \frac{1}{2^\rho} \sum_{V} \phi_V^\dagger \left(  D^2 + \frac12 \lambda_\text{ex} v^2 + \frac12 \lambda_V (v+h)^2 \right) \phi_V \,,
\label{eq:QuadraticScalarLagrangianDecomp}
\end{align}
where we have defined
\begin{equation}
\lambda_V = \lambda_{h\Phi} + \lambda^\prime_{h\Phi} \Big[ C_2(L) + C_2(R) - C_2(V) \Big] \,.
\end{equation}
Note that we only retain the strong and electromagnetic interactions in the covariant derivative $D$ in \cref{eq:QuadraticScalarLagrangianDecomp}; weak interactions generically couple different irreps of $\mathcal{V}$.

\cref{eq:QuadraticScalarLagrangianDecomp} generates one-loop corrections to the Higgs Lagrangian, which we calculate up to two derivative order using (D.21) of \cite{Cohen:2020xca}.
\begin{align}
\mathcal{L}_\text{eff}
&= \frac{1}{2^\rho (4\pi)^2} \sum_{V}\, V\, \Bigg\{
\frac12 \left[ \frac12 \lambda_\text{ex} v^2 + \frac12 \lambda_V (v+h)^2 \right]^2 \left[ \ln \frac{2\mu^2}{ \lambda_\text{ex} v^2 + \lambda_V (v+h)^2 } + \frac32 \right] \notag\\
&\hspace{100pt} + \frac{1}{24} \frac{\lambda_V^2}{ \lambda_\text{ex} v^2 + \lambda_V (v+h)^2} \left[ \partial (v+h)^2 \right]^2 + \mathcal{O}(\partial^4) \Bigg\} \,, \notag\\[5pt]
&\to \frac{1}{2^\rho (4 \pi)^2} \sum_{V}\, V\, \Bigg\{
\frac12 \left( \frac12 \lambda_\text{ex} v^2 + \lambda_V |H|^2 \right)^2 \left( \ln \frac{2\mu^2}{\lambda_\text{ex} v^2 + 2 \lambda_V |H|^2 } + \frac32 \right) \notag\\
&\hspace{100pt} + \frac13 \frac{\lambda_V^2}{\lambda_\text{ex} v^2 + 2 \lambda_V |H|^2} \frac{\left[ \partial |H|^2 \right]^2 }{2} \Bigg\}
+\mathcal{O}(\partial^4) \,,
\label{eq:ScalarOneLoop}
\end{align}
where we return to a general gauge by the substitution $(v+h)^2 \to 2 |H|^2$.

\subsection{Fermions}

The most general quadratic Lagrangian for a pair of Dirac fermions $\Psi_1,\Psi_2$, transforming under $[L_1,R_1]\equiv[2l_1+1,2r_1+1]$ and $[L_2,R_2]\equiv[2l_2+1,2r_2+1]$ respectively, is
\begin{equation}
\lag_\text{quad} = \tr\Big[ \overline{\Psi}_1 (i \slashed{D}-M_1) {\Psi_1} \Big] + \tr\Big[ \overline{\Psi}_2 (i \slashed{D}-M_2) {\Psi_2} \Big] - \left( \lag_\text{Yuk} + \hc \right) \,.
\label{eq:QuadraticFermionLagrangian}
\end{equation}
We assume w.l.o.g. that $M_1, M_2 > 0$. In the Yukawa term, one needs to contract the indices of fields properly to yield a custodial singlet
\begin{equation}
\lag_\text{Yuk} = y_{12} {\overline{\Psi}_1}_{\alpha\dot\alpha} H_{\beta\dot\beta} {\Psi_2}_{\gamma\dot\gamma}
\ip{\frac12 \beta ; l_2 \gamma }{ l_1 \alpha}\ip{r_1 \dot\alpha }{ \frac12 \dot\beta ;r_2 \dot\gamma} \,.
\label{eqn:lagYuk}
\end{equation}
Here we have shifted all the indices such that $\alpha$ runs in $\left[-l_1, l_1\right]$ (instead of $[1, 2l_1+1]$) and so on. This way we can identify the proper contraction coefficients as the standard Clebsch-Gordan coefficients, written in bra-ket notation.

As in the scalar case, we decompose fermionic matrix fields $\Psi_1, \Psi_2$ into their respective irreps under the diagonal subgroup $SU(2)_V \subset SU(2)_L \times SU(2)_R$:
\begin{equation}
\Psi_1 \to \underset{V_1}{\oplus}\, \psi_{1,V_1} \,,\qquad
\Psi_2 \to \underset{V_2}{\oplus}\, \psi_{2,V_2} \,,
\label{eqn:opluspsiV}
\end{equation}
where $\psi_{1,V_1}$($\psi_{2,V_2}$) are $V_1$($V_2$)-dimensional vectors, whose components are explicitly given by
\begin{subequations}\label{eqn:Psiinpsi}
\begin{align}
  {\Psi_1}_{\mu\dot\mu}  &= \sqrt{\frac{V_1}{L_1}}\ip{l_1 \mu }{ j_1 m_1 ; r_1 \dot\mu} \psi_{1,V_1}^{m_1} \,,\\[5pt]
  {\Psi_2}_{\nu\dot\nu}  &= \sqrt{\frac{V_2}{L_2}}\ip{l_2 \nu }{ j_2 m_2 ; r_2 \dot\nu} \psi_{2,V_2}^{m_2} \,.
\end{align}
\end{subequations}
In \cref{eqn:Psiinpsi}, the sums over the diagonal subgroup indices are
\begin{subequations}
\begin{align}
2 j_1 +1 = V_1\;\; &\in\;\; \mathcal{V}_1 = \Big\{L_1+R_1-1\;,\; L_1+R_1-3\;,\; \cdots \;,\; |L_1-R_1|+1 \Big\} \,,\notag\\[3pt]
&\text{with}\quad  -j_1 \leq m_1 \leq j_1 \,, \\[8pt]
2 j_2 +1 = V_2\;\; &\in\;\; \mathcal{V}_2 = \Big\{L_2+R_2-1\;,\; L_2+R_2-3\;,\; \cdots \;,\; |L_2-R_2|+1 \Big\} \,, \notag\\[3pt]
&\text{with}\quad -j_2 \leq m_2 \leq j_2  \,.
\end{align}
\end{subequations}

Using \cref{eqn:Psiinpsi}, we write the Yukawa piece of the Lagrangian \cref{eqn:lagYuk} in terms of $\psi_{1,V_1}$ and $\psi_{2,V_2}$, in unitary gauge ($H_{\beta\dot\beta} = \frac{1}{\sqrt{2}} (v+h) \delta_{\beta\dot\beta}$):
\begin{align}
  \lag_\text{Yuk} &= y_{12}\frac{1}{\sqrt{2}}(v+h) \, \overline{\psi}_{1,V_1}^{m_1} \psi_{2,V_2}^{m_2} \times \sqrt{\frac{V_1}{L_1}}\sqrt{\frac{V_2}{L_2}} \notag\\[5pt]
&\hspace{20pt}
\times \ip{l_1 \alpha }{ j_1 m_1 ; r_1 \dot\alpha } \ip{j_2 m_2 ; r_2 \dot\gamma }{ l_2 \gamma } \ip{\frac12 \beta; l_2 \gamma }{ l_1 \alpha}\ip{r_1 \dot\alpha }{ \frac12 \beta ;r_2 \dot\gamma} \, .
\end{align}
Summing over the Greek indices, the product of Clebsch-Gordan coefficients evaluates to a Wigner 6j symbol \cite[\textsection 12.1.4]{khersonskii1988quantum}
\begin{align}
 & \sqrt{\frac{V_1}{L_1}}\sqrt{\frac{V_2}{L_2}} \times \ip{l_1 \alpha }{ j_1 m_1 ; r_1 \dot\alpha } \ip{j_2 m_2 ; r_2 \dot\gamma }{ l_2 \gamma } \ip{\frac12 \beta; l_2 \gamma }{ l_1 \alpha}\ip{r_1 \dot\alpha }{ \frac12 \beta ;r_2 \dot\gamma} \notag \\[5pt]
 &= (-1)^{j_1+r_1+l_2+\frac12} \times \sqrt{L_1 R_1} \times \delta_{j_1,j_2} \delta_{m_1,m_2} \times
\left\{ \begin{matrix}
  r_2 & l_2 & j_1 \\ l_1 & r_1 & \frac12
\end{matrix}
\right\}
\end{align}
so the Lagrangian \cref{eq:QuadraticFermionLagrangian} decomposes as
\begin{align}
\lag_\text{quad} &= \sum_{V \in \mathcal{V}_1 - \mathcal{V}_2} \overline{\psi}_{1,V} \left( i\slashed{D} - M_1 \right) \psi_{1,V}
 + \sum_{V \in \mathcal{V}_2 - \mathcal{V}_1} \overline{\psi}_{2,V} \left( i\slashed{D} - M_2 \right) \psi_{2,V} \notag\\[5pt]
&+ \sum_{V \in \mathcal{V}_1 \cap \mathcal{V}_2} \mqty( \overline{\psi}_{1,V} & \overline{\psi}_{2,V} )
\left[ i\slashed{D} - \mqty( M_1 & \frac{1}{\sqrt{2}} y_V(v+h) \\ \frac{1}{\sqrt{2}} y_V^* (v+h) & M_2 ) \right]
\mqty(\psi_{1,V} \\ \psi_{2,V}) \,,
\label{eq:QuadraticFermionLagrangianDecomp}
\end{align}
where
\begin{equation}
  y_V = (-1)^{j_1+r_1+l_2+\frac12}\, y_{12}  \times \sqrt{L_1 R_1} \times
\left\{ \begin{matrix}
  r_2 & l_2 & j_1 \\ l_1 & r_1 & \frac12
\end{matrix}
\right\} \, .
\end{equation}

To integrate out the above quadratic piece, we use the machinery from \textsection6.3 of \cite{Cohen:2020xca}. Up to two derivative order
\begin{align}
\lag_\text{eff} &=
- \frac{1}{16\pi^2} \left[\, \sum_{V \in \mathcal{V}_1 - \mathcal{V}_2} V M_1^4 \left( \ln\frac{\mu^2}{M_1^2} + \frac32 \right)
+ \sum_{V \in \mathcal{V}_2 - \mathcal{V}_1} V M_2^4 \left( \ln\frac{\mu^2}{M_2^2} + \frac32 \right) \,\right]
\notag\\[5pt]
&\quad
- \frac{1}{16\pi^2} \sum_{V \in \mathcal{V}_1 \cap \mathcal{V}_2} V \Bigg\{\,
  \sum_{M=M_\pm} \left( M^4 + \frac12 |y_V|^2 (\partial h)^2\right) \left( \ln\frac{\mu^2}{M^2} + \frac32 \right)
\notag\\[5pt]
&\hspace{120pt}
+ \frac{4 |y_V|^4 (v+h)^2 (\partial h)^2}{3 (M_+ - M_-)^2}
\notag\\[5pt]
&\hspace{120pt}
+ M_+ M_- \bigg[ \frac{(M_+^2 + M_-^2)}{(M_+^2 - M_-^2)^2} - \frac{2 M_+^2 M_-^2}{(M_+^2 - M_-^2)^3} \ln\frac{M_+^2}{M_-^2} \bigg]
\notag\\[5pt]
&\hspace{150pt}
\times\bigg[ |y_V|^2 (\partial h)^2 - \frac{4 |y_V|^4 (v+h)^2 (\partial h)^2}{(M_+ - M_-)^2} \bigg]
\Bigg\} \,,
\label{eq:FermionOneLoop}
\end{align}
where
\begin{equation}
M_\pm = \frac12 (M_1 + M_2) \pm \frac12 \sqrt{(M_1-M_2)^2 + 2 |y_V|^2 (v+h)^2} \,.
\label{eq:MForFermionOneLoop}
\end{equation}

\addcontentsline{toc}{section}{\protect\numberline{}References}%
\bibliographystyle{JHEP}
\bibliography{Loryons}

\providecommand{\href}[2]{#2}\begingroup\raggedright\begin{thebibliography}{10}

\bibitem{Cohen:2020xca}
T.~Cohen, N.~Craig, X.~Lu and D.~Sutherland, \emph{{Is SMEFT Enough?}},
  \href{https://doi.org/10.1007/JHEP03(2021)237}{\emph{JHEP} {\bfseries 03}
  (2021) 237} [\href{https://arxiv.org/abs/2008.08597}{{\ttfamily
  2008.08597}}].

\bibitem{Alonso:2015fsp}
R.~Alonso, E.E.~Jenkins and A.V.~Manohar, \emph{{A Geometric Formulation of
  Higgs Effective Field Theory: Measuring the Curvature of Scalar Field
  Space}}, \href{https://doi.org/10.1016/j.physletb.2016.01.041}{\emph{Phys.
  Lett. B} {\bfseries 754} (2016) 335}
  [\href{https://arxiv.org/abs/1511.00724}{{\ttfamily 1511.00724}}].

\bibitem{Alonso:2016oah}
R.~Alonso, E.E.~Jenkins and A.V.~Manohar, \emph{{Geometry of the Scalar
  Sector}}, \href{https://doi.org/10.1007/JHEP08(2016)101}{\emph{JHEP}
  {\bfseries 08} (2016) 101}
  [\href{https://arxiv.org/abs/1605.03602}{{\ttfamily 1605.03602}}].

\bibitem{Falkowski:2019tft}
A.~Falkowski and R.~Rattazzi, \emph{{Which EFT}},
  \href{https://doi.org/10.1007/JHEP10(2019)255}{\emph{JHEP} {\bfseries 10}
  (2019) 255} [\href{https://arxiv.org/abs/1902.05936}{{\ttfamily
  1902.05936}}].

\bibitem{Cheung:2011aa}
C.~Cheung and Y.~Nomura, \emph{{Higgs Descendants}},
  \href{https://doi.org/10.1103/PhysRevD.86.015004}{\emph{Phys. Rev. D}
  {\bfseries 86} (2012) 015004}
  [\href{https://arxiv.org/abs/1112.3043}{{\ttfamily 1112.3043}}].

\bibitem{Djouadi:2012ae}
A.~Djouadi and A.~Lenz, \emph{{Sealing the fate of a fourth generation of
  fermions}}, \href{https://doi.org/10.1016/j.physletb.2012.07.060}{\emph{Phys.
  Lett. B} {\bfseries 715} (2012) 310}
  [\href{https://arxiv.org/abs/1204.1252}{{\ttfamily 1204.1252}}].

\bibitem{Kuflik:2012ai}
E.~Kuflik, Y.~Nir and T.~Volansky, \emph{{Implications of Higgs searches on the
  four generation standard model}},
  \href{https://doi.org/10.1103/PhysRevLett.110.091801}{\emph{Phys. Rev. Lett.}
  {\bfseries 110} (2013) 091801}
  [\href{https://arxiv.org/abs/1204.1975}{{\ttfamily 1204.1975}}].

\bibitem{Bonnefoy:2020gyh}
Q.~Bonnefoy, L.~Di~Luzio, C.~Grojean, A.~Paul and A.N.~Rossia, \emph{{The
  anomalous case of axion EFTs and massive chiral gauge fields}},
  \href{https://doi.org/10.1007/JHEP07(2021)189}{\emph{JHEP} {\bfseries 07}
  (2021) 189} [\href{https://arxiv.org/abs/2011.10025}{{\ttfamily
  2011.10025}}].

\bibitem{deBlas:2017xtg}
J.~de~Blas, J.C.~Criado, M.~Perez-Victoria and J.~Santiago, \emph{{Effective
  description of general extensions of the Standard Model: the complete
  tree-level dictionary}},
  \href{https://doi.org/10.1007/JHEP03(2018)109}{\emph{JHEP} {\bfseries 03}
  (2018) 109} [\href{https://arxiv.org/abs/1711.10391}{{\ttfamily
  1711.10391}}].

\bibitem{Cohen:2021ucp}
T.~Cohen, N.~Craig, X.~Lu and D.~Sutherland, \emph{{Unitarity Violation and the
  Geometry of Higgs EFTs}},  \href{https://arxiv.org/abs/2108.03240}{{\ttfamily
  2108.03240}}.

\bibitem{Jacob:1959at}
M.~Jacob and G.C.~Wick, \emph{{On the General Theory of Collisions for
  Particles with Spin}},
  \href{https://doi.org/10.1016/0003-4916(59)90051-X}{\emph{Annals Phys.}
  {\bfseries 7} (1959) 404}.

\bibitem{Goodsell:2018tti}
M.D.~Goodsell and F.~Staub, \emph{{Unitarity constraints on general scalar
  couplings with SARAH}},
  \href{https://doi.org/10.1140/epjc/s10052-018-6127-z}{\emph{Eur. Phys. J. C}
  {\bfseries 78} (2018) 649}
  [\href{https://arxiv.org/abs/1805.07306}{{\ttfamily 1805.07306}}].

\bibitem{Gunion:1989we}
J.F.~Gunion, H.E.~Haber, G.L.~Kane and S.~Dawson, \emph{{The Higgs Hunter's
  Guide}}, vol.~80 (2000).

\bibitem{Carmi:2012in}
D.~Carmi, A.~Falkowski, E.~Kuflik, T.~Volansky and J.~Zupan, \emph{{Higgs After
  the Discovery: A Status Report}},
  \href{https://doi.org/10.1007/JHEP10(2012)196}{\emph{JHEP} {\bfseries 10}
  (2012) 196} [\href{https://arxiv.org/abs/1207.1718}{{\ttfamily 1207.1718}}].

\bibitem{Aad:2019mbh}
{\scshape ATLAS} collaboration, \emph{{Combined measurements of Higgs boson
  production and decay using up to $80$ fb$^{-1}$ of proton-proton collision
  data at $\sqrt{s}=$ 13 TeV collected with the ATLAS experiment}},
  \href{https://doi.org/10.1103/PhysRevD.101.012002}{\emph{Phys. Rev. D}
  {\bfseries 101} (2020) 012002}
  [\href{https://arxiv.org/abs/1909.02845}{{\ttfamily 1909.02845}}].

\bibitem{Sirunyan:2018koj}
{\scshape CMS} collaboration, \emph{{Combined measurements of Higgs boson
  couplings in proton\textendash{}proton collisions at $\sqrt{s}=13\,\text
  {Te}\text {V} $}},
  \href{https://doi.org/10.1140/epjc/s10052-019-6909-y}{\emph{Eur. Phys. J. C}
  {\bfseries 79} (2019) 421}
  [\href{https://arxiv.org/abs/1809.10733}{{\ttfamily 1809.10733}}].

\bibitem{Lynn:1985fg}
B.W.~Lynn, M.E.~Peskin and R.G.~Stuart, \emph{{RADIATIVE CORRECTIONS IN SU(2) x
  U(1): LEP / SLC}},  7, 1985.

\bibitem{Peskin:1991sw}
M.E.~Peskin and T.~Takeuchi, \emph{{Estimation of oblique electroweak
  corrections}}, \href{https://doi.org/10.1103/PhysRevD.46.381}{\emph{Phys.
  Rev. D} {\bfseries 46} (1992) 381}.

\bibitem{Maksymyk:1993zm}
I.~Maksymyk, C.P.~Burgess and D.~London, \emph{{Beyond S, T and U}},
  \href{https://doi.org/10.1103/PhysRevD.50.529}{\emph{Phys. Rev. D} {\bfseries
  50} (1994) 529} [\href{https://arxiv.org/abs/hep-ph/9306267}{{\ttfamily
  hep-ph/9306267}}].

\bibitem{Burgess:1993mg}
C.P.~Burgess, S.~Godfrey, H.~Konig, D.~London and I.~Maksymyk, \emph{{A Global
  fit to extended oblique parameters}},
  \href{https://doi.org/10.1016/0370-2693(94)91322-6}{\emph{Phys. Lett. B}
  {\bfseries 326} (1994) 276}
  [\href{https://arxiv.org/abs/hep-ph/9307337}{{\ttfamily hep-ph/9307337}}].

\bibitem{Kundu:1996ah}
A.~Kundu and P.~Roy, \emph{{A General treatment of oblique parameters}},
  \href{https://doi.org/10.1142/S0217751X97001079}{\emph{Int. J. Mod. Phys. A}
  {\bfseries 12} (1997) 1511}
  [\href{https://arxiv.org/abs/hep-ph/9603323}{{\ttfamily hep-ph/9603323}}].

\bibitem{Barbieri:2004qk}
R.~Barbieri, A.~Pomarol, R.~Rattazzi and A.~Strumia, \emph{{Electroweak
  symmetry breaking after LEP-1 and LEP-2}},
  \href{https://doi.org/10.1016/j.nuclphysb.2004.10.014}{\emph{Nucl. Phys. B}
  {\bfseries 703} (2004) 127}
  [\href{https://arxiv.org/abs/hep-ph/0405040}{{\ttfamily hep-ph/0405040}}].

\bibitem{Farina:2016rws}
M.~Farina, G.~Panico, D.~Pappadopulo, J.T.~Ruderman, R.~Torre and A.~Wulzer,
  \emph{{Energy helps accuracy: electroweak precision tests at hadron
  colliders}},
  \href{https://doi.org/10.1016/j.physletb.2017.06.043}{\emph{Phys. Lett. B}
  {\bfseries 772} (2017) 210}
  [\href{https://arxiv.org/abs/1609.08157}{{\ttfamily 1609.08157}}].

\bibitem{ParticleDataGroup:2018ovx}
{\scshape Particle Data Group} collaboration, \emph{{Review of Particle
  Physics}}, \href{https://doi.org/10.1103/PhysRevD.98.030001}{\emph{Phys. Rev.
  D} {\bfseries 98} (2018) 030001}.

\bibitem{Khachatryan:2016sfv}
{\scshape CMS} collaboration, \emph{{Search for long-lived charged particles in
  proton-proton collisions at $\sqrt s=$ 13 TeV}},
  \href{https://doi.org/10.1103/PhysRevD.94.112004}{\emph{Phys. Rev. D}
  {\bfseries 94} (2016) 112004}
  [\href{https://arxiv.org/abs/1609.08382}{{\ttfamily 1609.08382}}].

\bibitem{Aaboud:2018hdl}
{\scshape ATLAS} collaboration, \emph{{Search for heavy charged long-lived
  particles in proton-proton collisions at $\sqrt{s} = 13$ TeV using an
  ionisation measurement with the ATLAS detector}},
  \href{https://doi.org/10.1016/j.physletb.2018.10.055}{\emph{Phys. Lett. B}
  {\bfseries 788} (2019) 96}
  [\href{https://arxiv.org/abs/1808.04095}{{\ttfamily 1808.04095}}].

\bibitem{Aaboud:2019trc}
{\scshape ATLAS} collaboration, \emph{{Search for heavy charged long-lived
  particles in the ATLAS detector in 36.1 fb$^{-1}$ of proton-proton collision
  data at $\sqrt{s} = 13$ TeV}},
  \href{https://doi.org/10.1103/PhysRevD.99.092007}{\emph{Phys. Rev. D}
  {\bfseries 99} (2019) 092007}
  [\href{https://arxiv.org/abs/1902.01636}{{\ttfamily 1902.01636}}].

\bibitem{Craig:2014lda}
N.~Craig, H.K.~Lou, M.~McCullough and A.~Thalapillil, \emph{{The Higgs Portal
  Above Threshold}}, \href{https://doi.org/10.1007/JHEP02(2016)127}{\emph{JHEP}
  {\bfseries 02} (2016) 127} [\href{https://arxiv.org/abs/1412.0258}{{\ttfamily
  1412.0258}}].

\bibitem{Crivellin:2020klg}
A.~Crivellin, F.~Kirk, C.A.~Manzari and L.~Panizzi, \emph{{Searching for lepton
  flavor universality violation and collider signals from a singly charged
  scalar singlet}},
  \href{https://doi.org/10.1103/PhysRevD.103.073002}{\emph{Phys. Rev. D}
  {\bfseries 103} (2021) 073002}
  [\href{https://arxiv.org/abs/2012.09845}{{\ttfamily 2012.09845}}].

\bibitem{Sirunyan:2018rlj}
{\scshape CMS} collaboration, \emph{{Search for pair-produced resonances
  decaying to quark pairs in proton-proton collisions at $\sqrt{s}=$ 13 TeV}},
  \href{https://doi.org/10.1103/PhysRevD.98.112014}{\emph{Phys. Rev. D}
  {\bfseries 98} (2018) 112014}
  [\href{https://arxiv.org/abs/1808.03124}{{\ttfamily 1808.03124}}].

\bibitem{Dercks:2018wch}
D.~Dercks and T.~Robens, \emph{{Constraining the Inert Doublet Model using
  Vector Boson Fusion}},
  \href{https://doi.org/10.1140/epjc/s10052-019-7436-6}{\emph{Eur. Phys. J. C}
  {\bfseries 79} (2019) 924}
  [\href{https://arxiv.org/abs/1812.07913}{{\ttfamily 1812.07913}}].

\bibitem{Bell:2020gug}
N.F.~Bell, M.J.~Dolan, L.S.~Friedrich, M.J.~Ramsey-Musolf and R.R.~Volkas,
  \emph{{Two-Step Electroweak Symmetry-Breaking: Theory Meets Experiment}},
  \href{https://doi.org/10.1007/JHEP05(2020)050}{\emph{JHEP} {\bfseries 05}
  (2020) 050} [\href{https://arxiv.org/abs/2001.05335}{{\ttfamily
  2001.05335}}].

\bibitem{Chiang:2020rcv}
C.-W.~Chiang, G.~Cottin, Y.~Du, K.~Fuyuto and M.J.~Ramsey-Musolf,
  \emph{{Collider Probes of Real Triplet Scalar Dark Matter}},
  \href{https://doi.org/10.1007/JHEP01(2021)198}{\emph{JHEP} {\bfseries 01}
  (2021) 198} [\href{https://arxiv.org/abs/2003.07867}{{\ttfamily
  2003.07867}}].

\bibitem{Abbiendi:2013hk}
{\scshape ALEPH, DELPHI, L3, OPAL, LEP} collaboration, \emph{{Search for
  Charged Higgs bosons: Combined Results Using LEP Data}},
  \href{https://doi.org/10.1140/epjc/s10052-013-2463-1}{\emph{Eur. Phys. J. C}
  {\bfseries 73} (2013) 2463}
  [\href{https://arxiv.org/abs/1301.6065}{{\ttfamily 1301.6065}}].

\bibitem{Aad:2021lzu}
{\scshape ATLAS} collaboration, \emph{{Search for doubly and singly charged
  Higgs bosons decaying into vector bosons in multi-lepton final states with
  the ATLAS detector using proton-proton collisions at $\sqrt{s}$ = 13 TeV}},
  \href{https://arxiv.org/abs/2101.11961}{{\ttfamily 2101.11961}}.

\bibitem{Aad:2019vnb}
{\scshape ATLAS} collaboration, \emph{{Search for electroweak production of
  charginos and sleptons decaying into final states with two leptons and
  missing transverse momentum in $\sqrt{s}=13$ TeV $pp$ collisions using the
  ATLAS detector}},
  \href{https://doi.org/10.1140/epjc/s10052-019-7594-6}{\emph{Eur. Phys. J. C}
  {\bfseries 80} (2020) 123}
  [\href{https://arxiv.org/abs/1908.08215}{{\ttfamily 1908.08215}}].

\bibitem{Aaboud:2017nhr}
{\scshape ATLAS} collaboration, \emph{{Search for the direct production of
  charginos and neutralinos in final states with tau leptons in $\sqrt{s} = $
  13 TeV $pp$ collisions with the ATLAS detector}},
  \href{https://doi.org/10.1140/epjc/s10052-018-5583-9}{\emph{Eur. Phys. J. C}
  {\bfseries 78} (2018) 154}
  [\href{https://arxiv.org/abs/1708.07875}{{\ttfamily 1708.07875}}].

\bibitem{Sirunyan:2017lae}
{\scshape CMS} collaboration, \emph{{Search for electroweak production of
  charginos and neutralinos in multilepton final states in proton-proton
  collisions at $\sqrt{s}=$ 13 TeV}},
  \href{https://doi.org/10.1007/JHEP03(2018)166}{\emph{JHEP} {\bfseries 03}
  (2018) 166} [\href{https://arxiv.org/abs/1709.05406}{{\ttfamily
  1709.05406}}].

\bibitem{Aaboud:2017mpt}
{\scshape ATLAS} collaboration, \emph{{Search for long-lived charginos based on
  a disappearing-track signature in pp collisions at $ \sqrt{s}=13 $ TeV with
  the ATLAS detector}},
  \href{https://doi.org/10.1007/JHEP06(2018)022}{\emph{JHEP} {\bfseries 06}
  (2018) 022} [\href{https://arxiv.org/abs/1712.02118}{{\ttfamily
  1712.02118}}].

\bibitem{Rentala:2011mr}
V.~Rentala, W.~Shepherd and S.~Su, \emph{{A Simplified Model Approach to
  Same-sign Dilepton Resonances}},
  \href{https://doi.org/10.1103/PhysRevD.84.035004}{\emph{Phys. Rev. D}
  {\bfseries 84} (2011) 035004}
  [\href{https://arxiv.org/abs/1105.1379}{{\ttfamily 1105.1379}}].

\bibitem{Aaboud:2017qph}
{\scshape ATLAS} collaboration, \emph{{Search for doubly charged Higgs boson
  production in multi-lepton final states with the ATLAS detector using
  proton\textendash{}proton collisions at $\sqrt{s}=13\,\text {TeV}$}},
  \href{https://doi.org/10.1140/epjc/s10052-018-5661-z}{\emph{Eur. Phys. J. C}
  {\bfseries 78} (2018) 199}
  [\href{https://arxiv.org/abs/1710.09748}{{\ttfamily 1710.09748}}].

\bibitem{Sirunyan:2020eab}
{\scshape CMS} collaboration, \emph{{Search for supersymmetry in final states
  with two oppositely charged same-flavor leptons and missing transverse
  momentum in proton-proton collisions at $\sqrt{s} =$ 13 TeV}},
  \href{https://doi.org/10.1007/JHEP04(2021)123}{\emph{JHEP} {\bfseries 04}
  (2021) 123} [\href{https://arxiv.org/abs/2012.08600}{{\ttfamily
  2012.08600}}].

\bibitem{Aad:2019vvi}
{\scshape ATLAS} collaboration, \emph{{Search for chargino-neutralino
  production with mass splittings near the electroweak scale in three-lepton
  final states in $\sqrt {s}$=13 TeV $pp$ collisions with the ATLAS detector}},
  \href{https://doi.org/10.1103/PhysRevD.101.072001}{\emph{Phys. Rev. D}
  {\bfseries 101} (2020) 072001}
  [\href{https://arxiv.org/abs/1912.08479}{{\ttfamily 1912.08479}}].

\bibitem{Aad:2019vvf}
{\scshape ATLAS} collaboration, \emph{{Search for direct production of
  electroweakinos in final states with one lepton, missing transverse momentum
  and a Higgs boson decaying into two $b$-jets in $pp$ collisions at
  $\sqrt{s}=13$ TeV with the ATLAS detector}},
  \href{https://doi.org/10.1140/epjc/s10052-020-8050-3}{\emph{Eur. Phys. J. C}
  {\bfseries 80} (2020) 691}
  [\href{https://arxiv.org/abs/1909.09226}{{\ttfamily 1909.09226}}].

\bibitem{CMS:2021lzg}
{\scshape CMS} collaboration, \emph{{Search for chargino-neutralino production
  in final states with a Higgs boson and a W boson}},  CMS-PAS-SUS-20-003,
  2021.

\bibitem{CMS:2019lwf}
{\scshape CMS} collaboration, \emph{{Search for physics beyond the standard
  model in multilepton final states in proton-proton collisions at $\sqrt{s} =$
  13 TeV}}, \href{https://doi.org/10.1007/JHEP03(2020)051}{\emph{JHEP}
  {\bfseries 03} (2020) 051}
  [\href{https://arxiv.org/abs/1911.04968}{{\ttfamily 1911.04968}}].

\bibitem{ATLAS:2021eyc}
{\scshape ATLAS} collaboration, \emph{{Search for new phenomena in three- or
  four-lepton events in $pp$ collisions at $\sqrt{s} = $ 13 TeV with the ATLAS
  detector}},  ATLAS-CONF-2021-011, 2021.

\bibitem{Thomas:1998wy}
S.D.~Thomas and J.D.~Wells, \emph{{Phenomenology of Massive Vectorlike Doublet
  Leptons}}, \href{https://doi.org/10.1103/PhysRevLett.81.34}{\emph{Phys. Rev.
  Lett.} {\bfseries 81} (1998) 34}
  [\href{https://arxiv.org/abs/hep-ph/9804359}{{\ttfamily hep-ph/9804359}}].

\bibitem{Cirelli:2005uq}
M.~Cirelli, N.~Fornengo and A.~Strumia, \emph{{Minimal dark matter}},
  \href{https://doi.org/10.1016/j.nuclphysb.2006.07.012}{\emph{Nucl. Phys. B}
  {\bfseries 753} (2006) 178}
  [\href{https://arxiv.org/abs/hep-ph/0512090}{{\ttfamily hep-ph/0512090}}].

\bibitem{ATLAS:2021ttq}
{\scshape ATLAS} collaboration, \emph{{Search for long-lived charginos based on
  a disappearing-track signature using 136 fb$^{-1}$ of $pp$ collisions at
  $\sqrt{s}$ = 13 TeV with the ATLAS detector}},  ATLAS-CONF-2021-015, 2021.

\bibitem{Alloul:2013bka}
A.~Alloul, N.D.~Christensen, C.~Degrande, C.~Duhr and B.~Fuks, \emph{{FeynRules
  2.0 - A complete toolbox for tree-level phenomenology}},
  \href{https://doi.org/10.1016/j.cpc.2014.04.012}{\emph{Comput. Phys. Commun.}
  {\bfseries 185} (2014) 2250}
  [\href{https://arxiv.org/abs/1310.1921}{{\ttfamily 1310.1921}}].

\bibitem{Hahn:2000kx}
T.~Hahn, \emph{{Generating Feynman diagrams and amplitudes with FeynArts 3}},
  \href{https://doi.org/10.1016/S0010-4655(01)00290-9}{\emph{Comput. Phys.
  Commun.} {\bfseries 140} (2001) 418}
  [\href{https://arxiv.org/abs/hep-ph/0012260}{{\ttfamily hep-ph/0012260}}].

\bibitem{Hahn:1998yk}
T.~Hahn and M.~Perez-Victoria, \emph{{Automatized one loop calculations in
  four-dimensions and D-dimensions}},
  \href{https://doi.org/10.1016/S0010-4655(98)00173-8}{\emph{Comput. Phys.
  Commun.} {\bfseries 118} (1999) 153}
  [\href{https://arxiv.org/abs/hep-ph/9807565}{{\ttfamily hep-ph/9807565}}].

\bibitem{Hahn:2016ebn}
T.~Hahn, S.~Pa\ss{}ehr and C.~Schappacher, \emph{{FormCalc 9 and Extensions}},
  \href{https://doi.org/10.1088/1742-6596/762/1/012065}{\emph{PoS} {\bfseries
  LL2016} (2016) 068} [\href{https://arxiv.org/abs/1604.04611}{{\ttfamily
  1604.04611}}].

\bibitem{lepsusy}
{\scshape LEPSUSYWG, ALEPH, DELPHI, L3 and OPAL} collaboration, \emph{{Combined
  lep chargino results, up to 208 gev for low dm}},
  \url{http://lepsusy.web.cern.ch/lepsusy/www/inoslowdmsummer02/charginolowdm_pub.html},
  2002.

\bibitem{Egana-Ugrinovic:2018roi}
D.~Egana-Ugrinovic, M.~Low and J.T.~Ruderman, \emph{{Charged Fermions Below 100
  GeV}}, \href{https://doi.org/10.1007/JHEP05(2018)012}{\emph{JHEP} {\bfseries
  05} (2018) 012} [\href{https://arxiv.org/abs/1801.05432}{{\ttfamily
  1801.05432}}].

\bibitem{Cepeda:2019klc}
M.~Cepeda et~al., \emph{{Report from Working Group 2}: {Higgs Physics at the
  HL-LHC and HE-LHC}},
  \href{https://doi.org/10.23731/CYRM-2019-007.221}{\emph{CERN Yellow Rep.
  Monogr.} {\bfseries 7} (2019) 221}
  [\href{https://arxiv.org/abs/1902.00134}{{\ttfamily 1902.00134}}].

\bibitem{Silveira:1985rk}
V.~Silveira and A.~Zee, \emph{{Scalar Phantoms}},
  \href{https://doi.org/10.1016/0370-2693(85)90624-0}{\emph{Phys. Lett. B}
  {\bfseries 161} (1985) 136}.

\bibitem{McDonald:1993ex}
J.~McDonald, \emph{{Gauge singlet scalars as cold dark matter}},
  \href{https://doi.org/10.1103/PhysRevD.50.3637}{\emph{Phys. Rev. D}
  {\bfseries 50} (1994) 3637}
  [\href{https://arxiv.org/abs/hep-ph/0702143}{{\ttfamily hep-ph/0702143}}].

\bibitem{Burgess:2000yq}
C.P.~Burgess, M.~Pospelov and T.~ter Veldhuis, \emph{{The Minimal model of
  nonbaryonic dark matter: A Singlet scalar}},
  \href{https://doi.org/10.1016/S0550-3213(01)00513-2}{\emph{Nucl. Phys. B}
  {\bfseries 619} (2001) 709}
  [\href{https://arxiv.org/abs/hep-ph/0011335}{{\ttfamily hep-ph/0011335}}].

\bibitem{Cohen:2011ec}
T.~Cohen, J.~Kearney, A.~Pierce and D.~Tucker-Smith, \emph{{Singlet-Doublet
  Dark Matter}}, \href{https://doi.org/10.1103/PhysRevD.85.075003}{\emph{Phys.
  Rev. D} {\bfseries 85} (2012) 075003}
  [\href{https://arxiv.org/abs/1109.2604}{{\ttfamily 1109.2604}}].

\bibitem{khersonskii1988quantum}
V.~Khersonskii, A.~Moskalev and D.~Varshalovich, \emph{Quantum Theory Of
  Angular Momemtum}, World Scientific Publishing Company (1988).

\end{thebibliography}\endgroup

\end{document}